\documentclass[10pt,aps,prx,superscriptaddress,nofootinbib,longbibliography]{revtex4-2}
\usepackage{amsmath,amssymb,amsfonts}
\usepackage{graphicx}

\usepackage{xcolor}

\usepackage{amsthm}

\usepackage{extarrows}

\usepackage[most]{tcolorbox}
\tcbuselibrary{breakable,skins}
\usepackage{xparse} 

\newtcolorbox{naturethmbox}[1][]{%
  enhanced,
  breakable,
  sharp corners,
  boxrule=0pt,
  frame hidden,
  colback=black!4.5,
  borderline west={0.7pt}{0pt}{black!50},
  left=7pt,
  right=7pt,
  top=5pt,
  bottom=5pt,
  before skip=8pt,
  after skip=8pt,
  #1
}

\newtcolorbox{naturedefbox}[1][]{%
  enhanced,
  breakable,
  sharp corners,
  boxrule=0pt,
  frame hidden,
  colback=black!4.5,
  borderline west={0.6pt}{0pt}{black!40},
  left=7pt,
  right=7pt,
  top=5pt,
  bottom=5pt,
  before skip=8pt,
  after skip=8pt,
  #1
}

\NewDocumentCommand{\bmyt}{o}{%
  \begin{naturethmbox}%
  \IfNoValueTF{#1}{\begin{theorem}}{\begin{theorem}[#1]}%
}
\newcommand{\emyt}{%
  \end{theorem}%
  \end{naturethmbox}%
}

\NewDocumentCommand{\bmyl}{o}{%
  \begin{naturethmbox}%
  \IfNoValueTF{#1}{\begin{lemma}}{\begin{lemma}[#1]}%
}
\newcommand{\emyl}{%
  \end{lemma}%
  \end{naturethmbox}%
}

\NewDocumentCommand{\bmyd}{o}{%
  \begin{naturedefbox}%
  \IfNoValueTF{#1}{\begin{definition}}{\begin{definition}[#1]}%
}
\newcommand{\emyd}{%
  \end{definition}%
  \end{naturedefbox}%
}

\NewDocumentCommand{\bmyc}{o}{%
  \begin{naturethmbox}%
  \IfNoValueTF{#1}{\begin{corollary}}{\begin{corollary}[#1]}%
}
\newcommand{\emyc}{%
  \end{corollary}%
  \end{naturethmbox}%
}

\usepackage{dsfont}

\usepackage{mathtools}

\def\1{\mathbf{1}}

\def\comp{{\rm Comp}}

\def\cptp{{\rm CPTP}}

\def\mi{{\mathfrak{I}}}

\newcommand{\epm}{\end{pmatrix}}
\newcommand{\bpm}{\begin{pmatrix}}

\renewcommand{\log}{{\operatorname{log}}}

\newcommand{\ebm}{\end{bmatrix}}
\newcommand{\bbm}{\begin{bmatrix}}

\def\>{\rangle}
\def\<{\langle}

\def\id{\mathsf{id}}

\def\mE{\mathcal{E}}

\def\mN{\mathcal{N}}

\def\mP{\mathcal{P}}

\def\mD{\mathcal{D}}
\def\mT{\mathcal{T}}

\def\mU{\mathcal{U}}

\newcommand{\supp}{\operatorname{supp}}

\renewcommand{\ge}{\geqslant}

\renewcommand{\geq}{\geqslant}
\renewcommand{\leq}{\leqslant}

\newcommand{\ben}{\begin{enumerate}}
\newcommand{\een}{\end{enumerate}}

\theoremstyle{definition}
\newtheorem{theorem}{Theorem}
\newtheorem{lemma}[theorem]{Lemma}
\newtheorem{corollary}[theorem]{Corollary}

\newtheorem{definition}[theorem]{Definition}

\usepackage{array}

\newcommand{\bea}{\begin{eqnarray}}
\newcommand{\eea}{\end{eqnarray}}
\newcommand{\be}{\begin{equation}}
\newcommand{\ee}{\end{equation}}
\newcommand{\ba}{\begin{equation}\begin{aligned}}
\newcommand{\ea}{\end{aligned}\end{equation}}

\newcommand{\bee}{\begin{enumerate}}
\newcommand{\eee}{\end{enumerate}}

\newcommand{\rank}{{\rm Rank}}

\newcommand{\ml}{\mathfrak{L}}
\newcommand{\mb}{\mathfrak{B}}
\newcommand{\md}{\mathfrak{D}}
\newcommand{\mf}{\mathfrak{F}}

\newcommand{\mfu}{\mathfrak{U}}

\newcommand{\mM}{\mathcal{M}}

\newcommand{\lr}{\rangle\langle}
\newcommand{\la}{\langle}
\newcommand{\ra}{\rangle}
\newcommand{\tr}{{\rm Tr}}

\newcommand{\eps}{\varepsilon}

\newcommand{\eqdef}{\coloneqq}

\def\br{\mathbf{r}}

\def\p{\mathbf{p}}
\def\q{\mathbf{q}}

\def\bt{\mathbf{t}}

\def\0{\mathbf{0}}

\def\trho{{\tilde{\rho}}}

\usepackage[hidelinks]{hyperref}
\hypersetup{
  pdftitle={Exact Quantum Compression with and without Classical Communication},
  pdfauthor={Gilad Gour}
}

\begin{document}

\title{Classical Communication Helps Single-Shot Quantum Compression}
\author{Omri Farber}
\affiliation{Faculty of Mathematics, Helen Diller Quantum Center, Technion--Israel Institute of Technology, Haifa 3200003, Israel}
\author{Gilad Gour}
\affiliation{Faculty of Mathematics, Helen Diller Quantum Center, Technion--Israel Institute of Technology, Haifa 3200003, Israel}
\affiliation{Mathematics with Computer Science Program, Guangdong Technion–Israel Institute of Technology, Shantou, Guangdong 515063, China}

\begin{abstract}
Free classical communication can yield an unbounded
single-shot saving in quantum compression, even though
it leaves the asymptotic rate of an i.i.d. source unchanged.
At every fixed error in $(0,1)$, we exhibit sources whose
unassisted compression cost is arbitrarily large, yet
a single qubit suffices with classical communication
and no preshared entanglement.
We establish this separation by determining the exact
optimal source fidelity for compression through an
$m$-dimensional noiseless quantum channel, with and
without free classical communication, while preserving
correlations with an inaccessible reference.
Without classical assistance, the optimal fidelity is
exactly the Ky Fan norm $\|\rho\|_{(m)}$, the sum of the
$m$ largest eigenvalues of the source state.
With classical assistance, we obtain an explicit spectral
threshold formula and a protocol attaining it.
Classical communication strictly improves fidelity whenever
$m$ is smaller than the source rank.
The corresponding compression costs are two smoothed
versions of the log-rank entropy, also called max-entropy:
trace-distance smoothing under local operations and
quadratic smoothing under LOCC.
For i.i.d. copies of a fixed source with positive entropy
variance, however, the two costs have identical first-,
second-, and third-order asymptotic terms.
For nonlattice spectra, we determine the constant-order
advantage: classical communication saves $\log_2(e)-1$
bits per block asymptotically, and the ratio of assisted
to unassisted minimum communication dimensions converges
to $2/e$.
\end{abstract}

\maketitle

\section{Introduction}\label{sec:introduction}

Quantum compression determines how much quantum communication
is needed to transfer a source faithfully~\cite{S1995}.
In the standard entanglement-fidelity formulation, faithful
transfer requires preserving the source's correlations with
an inaccessible reference, or equivalently preserving a
purification of the source~\cite{S1996}. The communication
cost depends on the resources available to the encoder and
decoder. Here we investigate how free classical communication
changes this cost for a finite source.

For independent and identically distributed sources, the
asymptotic compression rate is the von Neumann entropy,
whether or not classical communication is
free~\cite{S1995,W1999,Sh2014}. How much classical messages
can help at a fixed accuracy is a different question.
We show that their benefit can be unbounded: at every
fixed fidelity strictly between zero and one, there are
sources whose unassisted compression cost is arbitrarily
large, yet a single qubit suffices with free classical
communication. Equality of asymptotic rates therefore
conceals a substantial difference in the resources
required for a finite source.

We determine this difference by solving the fidelity
optimization exactly in both models. The communication
resource is an $m$-dimensional noiseless quantum channel
$\id_m$, whose cost is $\log(m)$, with logarithms base two.
Under local operations (LO), Alice encodes and Bob decodes
using arbitrary local channels. Under local operations
and classical communication (LOCC), they may additionally
exchange classical messages in either direction.
No preshared entanglement is available, and Alice receives
the quantum input without a preparation label.

The accuracy criterion distinguishes this task from other
quantum--classical compression tradeoffs.
Visible compression provides the preparation label to
the encoder~\cite{HJW2002}, while rate-distortion coding
allows reconstruction according to a specified distortion
criterion~\cite{DB2002,WDHW2013,DRRW2013}.
Here we require fidelity to the entire purified source.
For a source state $\rho$ on $A$, choose a purification
$\psi_{RA}$, and let $B$ be a copy of $A$.
The source fidelity achieved by a channel $\mN_{A\to B}$ is
\be\label{intro:source-fidelity}
F_\rho(\mN,\id_{A\to B})
\eqdef
\sqrt{\la\psi_{RB}|
\mN_{A\to B}(\psi_{RA})|\psi_{RB}\ra},
\ee
where the channel acts implicitly as the identity on $R$.
Its square is the standard entanglement
fidelity~\cite{S1996}.
We write
$F_\rho(\id_m\xrightarrow{\mf}\id_{A\to B})$
for the optimum over protocols using $\id_m$ and
operations $\mf\in\{{\rm LO},{\rm LOCC}\}$.

Our main characterizations are
\ba\label{intro:optimal-fidelities}
F_\rho\left(\id_m\xrightarrow{\rm LO}\id_{A\to B}\right)
&=\|\rho\|_{(m)},\\
F_\rho\left(\id_m\xrightarrow{\rm LOCC}\id_{A\to B}\right)
&=\max_{\substack{0\leq\Lambda\leq I\\
\tr[\Lambda^2]\leq m}}\tr[\rho\Lambda].
\ea
The Ky Fan norm $\|\rho\|_{(m)}$ is the sum of the
$m$ largest eigenvalues of $\rho$.
The simplicity of the LO identity is notable:
an optimization over all encoding and decoding channels
reduces to a sum of eigenvalues. Spectral truncation
attains this value, and our converse establishes that
it is optimal at every communication dimension.

Classical communication replaces this single-subspace
optimization by a quadratic effect optimization.
Operationally, Alice applies an encoding instrument
and sends its outcome to Bob, allowing him to select
the corresponding decoding isometry.
We construct an instrument attaining the quadratic
optimum and prove a matching converse for general
LOCC protocols. The optimum also admits an explicit
spectral threshold formula, given in \eqref{loccexact}.
Classical communication strictly improves fidelity
whenever $m<\operatorname{rank}\rho$.

The quadratic program, in its inverse formulation,
and its clipped optimizer were previously obtained for
partially smoothed conditional min-entropy of a pure
state~\cite[Sec.~III]{AR2020}.
The associated spectral norm also appears in one-shot
coherence and entanglement
distillation~\cite{RFWA2018,RFWG2019}.
Our contribution is the exact compression interpretation:
we show that this value both bounds every admissible
protocol and is attainable while preserving the
inaccessible reference.

To express the resulting saving in communication cost,
let $\comp_{\mf}^\eps(\rho)$ denote the minimum $\log(m)$
required to attain source fidelity at least $1-\eps$.
For every fixed $\eps\in(0,1)$, we exhibit sources
$\rho_n$ of rank $n^2+1$, for arbitrarily large integers
$n$, satisfying
\ba\label{intro:unbounded-saving}
\comp_{\rm LO}^\eps(\rho_n)&=\log(n+1),\\
\comp_{\rm LOCC}^\eps(\rho_n)&=1.
\ea
Thus the same reconstruction accuracy can require an
arbitrarily large quantum dimension under LO and
dimension two under LOCC. This separation requires
neither access to a preparation label nor a relaxation
to symbolwise reconstruction. It arises from classical
coordination of the encoder and decoder under the
purification criterion itself.

The exact formulas also identify the two costs
entropically:
\ba\label{intro:compression-costs}
\comp_{\rm LO}^\eps(\rho)
&=H_{\max}^\eps(A)_\rho,\\
\comp_{\rm LOCC}^\eps(\rho)
&=H_{\max}^{\eps,{\rm q}}(A)_\rho.
\ea
These are respectively the standard trace-distance
smoothing and a quadratic smoothing of the log-rank
entropy, also called max-entropy.
The first selects a subspace carrying sufficient source
weight~\cite{TFG2025}; the second allows an effect and
charges its squared Hilbert--Schmidt norm, rounded up
to an integer dimension.
Both reduce to $\log|\supp(\rho)|$ at zero error.
Their definitions and spectral expressions are given
in Section~\ref{sec:entropies}.

The unbounded separation concerns a family of changing
sources. Repeated copies of one fixed source exhibit
a different behavior.
One-shot entropic bounds for blind compression led
to second-order bounds~\cite{DL2015}, whose gap was
closed using the partial-smoothing state-merging
converse~\cite{ABJT2020,AR2020}.
Together with the classical source-coding
expansions~\cite{KV2014,H2018}, these methods account
for the common first-, second-, and third-order terms
of the LO and LOCC costs.
We resolve their remaining constant-order difference.
For a fixed source with positive entropy variance,
whose positive log-eigenvalues do not all lie on an
arithmetic lattice, and every fixed $\eps\in(0,1)$,
\be\label{intro:asymptotic-saving}
\lim_{n\to\infty}\left(
\comp_{\rm LO}^\eps(\rho^{\otimes n})
-\comp_{\rm LOCC}^\eps(\rho^{\otimes n})
\right)
=\log_2(e)-1.
\ee
Equivalently, the ratio of the minimum assisted to
unassisted communication dimensions converges to $2/e$.
The saving per source copy vanishes, while the reduction
in the dimension needed for the whole block remains.

These results distinguish two ways in which classical
communication matters beyond an entropy rate:
it can keep the quantum cost bounded across a family
with diverging unassisted cost, and it can produce a
definite multiplicative reduction in communication
dimension for repeated copies of a fixed source.
The exact fidelity formulas make both effects accessible
without an optimization over encoding and decoding
channels. Their connection to partial smoothing also
motivates a broader question: whether similarly exact
operational characterizations can be obtained for
quantum state merging, where the receiver already
holds quantum side information.

Section~\ref{sec:preliminaries} specifies the task and
notation, and Section~\ref{sec:entropies} introduces
the two smoothings.
Sections~\ref{sec:lo} and~\ref{sec:locc} establish
the exact compression characterizations.
Section~\ref{sec:asymptotics} compares the i.i.d.
costs, and Appendix~\ref{app:smoothing} discusses
extensions of operator smoothing.

\section{Preliminaries}\label{sec:preliminaries}

This section collects the notation, standard fidelity formulas,
and operational definitions used throughout the paper. We specify
the source and its reference, the allowed operations, and the
communication cost.
All systems are finite dimensional. We write $\md(A)$ for the
density operators on $A$ and $\cptp(A\to B)$ for quantum channels. The identity
operator is $I$, and the identity channel on an $m$-dimensional
system is $\id_m$. All logarithms are base two. We use $[d]\eqdef\{1,\ldots,d\}$,
$\|\cdot\|_p$ for the Schatten norms, and $\mfu(d)$ for the
unitary group. We denote the trace-distance ball of radius $\eps\in[0,1]$
around $\rho$ by
\be\label{trace-distance-ball}
\mb^\eps(\rho)\eqdef
\left\{\trho\in\md(A)\;:\;
\frac12\|\trho-\rho\|_1\leq\eps\right\}.
\ee

For a nonnegative vector $\p\in\mathbb R_+^d$, we write
$\|\p\|_{(m)}\eqdef\sum_{x=1}^m p_x^\downarrow$
for the sum of its $m$ largest components, where
$\p^\downarrow$ denotes its nonincreasing rearrangement.
For $\p,\q\in\mathbb R_+^d$ with equal total sums, we say
that $\p$ is majorized by $\q$, denoted by $\p\prec\q$, if
\be
\|\p\|_{(m)}\leq\|\q\|_{(m)}
\qquad\forall\,m\in[d-1].
\ee
For $\rho\in\md(A)$, let $\lambda(\rho)$ be its eigenvalue
vector in nonincreasing order. The same notation gives
$\|\rho\|_{(m)}\eqdef\|\lambda(\rho)\|_{(m)}$.
In particular, for $\rho,\sigma\in\md(A)$,
\be
\lambda(\rho)\prec\lambda(\sigma)
\quad\Longleftrightarrow\quad
\|\rho\|_{(m)}\leq\|\sigma\|_{(m)}
\qquad\forall\,m\in[d-1].
\ee

\subsection{A source and its inaccessible reference}

Let $\rho_A$ be the source state and $\psi_{RA}$ a
purification, so that $\tr_R[\psi_{RA}]=\rho_A$.
Alice must transfer $A$ to Bob while preserving its
correlations with the inaccessible reference $R$.
Preserving this purification also ensures preservation
of every extension of $\rho_A$, whereas reproducing
$\rho_A$ alone would require no communication.
The information that must be preserved is therefore
part of the operational specification: other compression
models preserve specified measurement
statistics~\cite{BRW2018} or a given bipartite
state~\cite{K2025}.
Our criterion also differs from the ensemble-average
reconstruction fidelity used in quantum autoencoder
optimization~\cite{CPL2026}, which tests a specified
pure-state ensemble without requiring preservation
of its coherent purification.
For an independent and identically distributed source,
the joint input is $\psi_{RA}^{\otimes n}$.

To quantify the accuracy of this preservation, let $B$ be a
copy of $A$ and put $d\eqdef|A|$. For positive operators,
we use the fidelity convention
$F(\omega,\tau)\eqdef\|\sqrt{\omega}\sqrt{\tau}\|_1$.
Given a purification $\psi_{RA}$ of $\rho_A$, the source
fidelity of channels $\mN,\mM\in\cptp(A\to B)$ is
\be\label{sfidelity}
F_\rho(\mN,\mM)
\eqdef
F\big(
\mN_{A\to B}(\psi_{RA}),
\mM_{A\to B}(\psi_{RA})
\big),
\ee
where channels implicitly act as the identity on $R$.
For the identity target, the squared source fidelity is
the entanglement fidelity~\cite{S1996}.
Writing $\psi_{RB}$ for the target purification with $A$
relabeled as $B$, and identifying $B$ with $A$ when taking
traces, any Kraus representation $\{M_x\}_x$ of $\mN$ gives
\ba\label{ex14}
F_\rho^2(\mN,\id_{A\to B})
&=\la\psi_{RB}|
\mN_{A\to B}(\psi_{RA})
|\psi_{RB}\ra\\
&=\sum_x\big|\tr[M_x\rho_A]\big|^2.
\ea
The first equality follows because the target purification
is pure, while the second is the standard Kraus expression
for entanglement fidelity~\cite{S1996}. We will use this
expression to optimize over encoders and decoders.  Exact transfer
is equivalent to the channel acting as the identity on all
operators supported on $\supp(\rho_A)$. Allowing an error
in the source fidelity relaxes this requirement and can
permit transmission through a smaller quantum system.

\subsection{Free operations and communication cost}

Having specified the accuracy criterion, we now describe the
operations available to Alice and Bob. Under local operations
(LO), Alice applies an encoding channel before transmission,
and Bob applies a decoding channel afterward. Their processing
transforms a communication resource $\mN\in\cptp(A'\to B')$ according to
\be\label{losc}
\Theta[\mN]
=\mD_{B'\to B}\circ\mN_{A'\to B'}\circ\mE_{A\to A'}.
\ee
Both channels are arbitrary and may use local ancillary systems.
They are chosen in advance, so Bob's decoding cannot depend on
an outcome observed by Alice.

Under local operations and classical communication (LOCC), Alice
and Bob may exchange classical messages freely before and after
transmission. A general LOCC superchannel has the form
\be\label{locc-general}
\Theta[\mN]
=\mD_{A''B'\to B}\circ\mN_{A'\to B'}\circ
\mE_{A\to A''A'},
\ee
where $\mE$ is a local encoding channel, $A''$ is a quantum
memory retained by Alice, and $\mD$ is an LOCC channel acting
on Alice's $A''$ and Bob's $B'$. As usual, the system labels
specify where each channel acts. Classical exchanges before
transmission can be included by storing their record in $A''$
and preparing any conditional ancillary state at Bob's laboratory
during postprocessing. Neither LO nor LOCC permits preshared
entanglement or additional quantum communication.

We take the communication resource to be the noiseless channel
$\id_m$ on an $m$-dimensional system. For either class of
operations $\mf\in\{{\rm LO},{\rm LOCC}\}$, its optimal fidelity for transmitting the source
is obtained by maximizing over the allowed processing:
\be\label{conversion-fidelity}
F_\rho\left(\id_m\xrightarrow{\mf}\id_{A\to B}\right)
\eqdef\max_{\Theta\in\mf}
F_\rho\big(\Theta[\id_m],\id_{A\to B}\big)
\,.
\ee
This quantity describes the best accuracy at a fixed quantum
dimension. Conversely, fixing the required fidelity determines
the smallest dimension that suffices. Measuring quantum
communication by $\log (m)$, with logarithms base two, gives
the following compression cost.

\bmyd\label{def:compression}
Let $\rho\in\md(A)$, $\eps\in[0,1)$, $d\eqdef|A|$, and
$\mf\in\{{\rm LO},{\rm LOCC}\}$. The $\eps$-single-shot
compression cost under $\mf$ is
\be\label{defc}
\comp_{\mf}^{\eps}(\rho)
\eqdef\min_{m\in[d]}
\left\{\log (m)\;:\;
F_\rho\left(\id_m\xrightarrow{\mf}\id_{A\to B}\right)
\geq1-\eps\right\}.
\ee
\emyd

The error parameter $\eps$ refers to source fidelity.
Since entanglement fidelity is its square, this definition
allows entanglement infidelity at most $2\eps-\eps^2$.
An entanglement-fidelity threshold $1-\delta$ therefore
corresponds to $\eps=1-\sqrt{1-\delta}$. We use the source
fidelity convention throughout, including in the asymptotic
results. At the boundary $\eps=1$, dimension one always
suffices, so both compression costs are zero.

\section{Two Smoothed Compression Entropies}\label{sec:entropies}

We introduce two smooth variants of the order-zero R\'enyi
entropy, which we call the max-entropy and denote by
\be
H_{\max}(A)_\rho\eqdef\log|\supp(\rho)|\,.
\ee
The first is
the standard trace-distance smoothing of the max entropy~\cite{RW2004,DR2009}
\be\label{smooth-max-definition}
H_{\max}^\eps(A)_\rho
\eqdef\min_{\trho\in\mb^\eps(\rho)}
H_{\max}(A)_{\trho}.
\ee
This is the order-zero smooth R\'enyi entropy~\cite{RW2004,DR2009,G2025};
throughout the paper, we use $H_{\max}^\eps$ to denote smoothed
log-rank. It is the logarithm of the smallest support dimension
compatible with the prescribed error. The spectral characterization
in~\cite[Corollary~2]{TFG2025} connects this state approximation problem to
the selection of a subspace carrying sufficient source weight:
\ba\label{smooth-max-spectral}
H_{\max}^\eps(A)_\rho
&=\log\min
\left\{m\in[d]\;:\;\|\rho\|_{(m)}\geq1-\eps\right\}\\
&=\log\min\left\{\tr[\Pi]\;:\;\Pi=\Pi^*=\Pi^2\,,\quad\tr[\rho\Pi]\geq1-\eps\right\}\,.
\ea
The second equality
follows from the variational principle for the Ky Fan norm:
among rank-$m$ projectors, the largest value of $\tr[\rho\Pi]$
is $\|\rho\|_{(m)}$, attained by projecting onto eigenvectors
associated with the $m$ largest eigenvalues of $\rho$.

The projector formulation identifies the information retained
by a subspace with the weight $\tr[\rho\Pi]$ it accepts.
Its dimension can equally be written as $\tr[\Pi^2]$.
This gives a natural relaxation: allow each spectral direction
to carry a weight between zero and one, by replacing the
projector with an effect, while keeping the quadratic cost.
This relaxation leads to a second smoothing of log-rank.
We denote it by $H_{\max}^{\eps,{\rm q}}$, where ${\rm q}$
indicates quadratic smoothing, and retain $H_{\max}^\eps$
for trace-distance smoothing. Both recover log-rank at
zero error.

\bmyd\label{def:quadratic} 
Let $\eps\in[0,1)$ and $\rho\in\md(A)$. The quadratically $\eps$-smoothed log-rank entropy of $\rho$
is
\be\label{quadratic-entropy}
H_{\max}^{\eps,{\rm q}}(A)_\rho
\eqdef\log\Bigl\lceil\min_\Lambda\tr\left[\Lambda^2\right]\Bigr\rceil,
\ee
where the minimum is over all effects $0\leq \Lambda\leq I$, satisfying
$\tr[\rho \Lambda]\geq1-\eps$.
\emyd

The ceiling converts the quadratic size of an effect into
an integer dimension. Since $\eps<1$, the minimum is strictly
positive, so $H_{\max}^{\eps,{\rm q}}(A)_\rho$ is nonnegative.
At zero error, every feasible effect acts as the identity
on $\supp(\rho)$, giving
$H_{\max}^{0,{\rm q}}(A)_\rho=\log|\supp(\rho)|$.

The optimization inside the ceiling is the spectral quadratic
program studied in~\cite[Sec.~III]{AR2020} through the partially
smoothed conditional min-entropy of a pure state. Their
purified-distance parameter is $\sqrt{1-(1-\eps)^2}$ in our
convention. Our definition rounds the optimal quadratic cost
up to an integer dimension before taking its logarithm.
The following properties connect this quantity to smoothed
log-rank and describe its behavior under mixing.

\bmyl\label{thm:quadratic-properties}
For every $\eps\in[0,1)$, $H_{\max}^{\eps,{\rm q}}$ is Schur-concave:
\be\label{quadratic-schur}
\lambda(\rho)\prec\lambda(\sigma)
\quad\Longrightarrow\quad
H_{\max}^{\eps,{\rm q}}(A)_\rho\geq H_{\max}^{\eps,{\rm q}}(A)_\sigma\,.
\ee
Moreover, for every $\rho\in\md(A)$
\be\label{entropy-comparison}
0\leq H_{\max}^{\eps,{\rm q}}(A)_\rho
\leq H_{\max}^\eps(A)_\rho\,.
\ee
Both smoothed entropies reduce to $\log|\supp(\rho)|$ at
zero error, are nonincreasing in $\eps$, and vanish for
pure states.
\emyl

\begin{proof}
To prove Schur concavity, we convert every feasible effect
for $\rho$ into one for $\sigma$ without increasing its
quadratic cost.
Since $\lambda(\rho)\prec\lambda(\sigma)$, there are
probabilities $t_z$ and unitaries $U_z$ such that
$\rho=\sum_z t_zU_z\sigma U_z^*$.
For a feasible effect $\Lambda$, define
$\widetilde\Lambda\eqdef\sum_z t_zU_z^*\Lambda U_z$.
This is an effect, and
\be
\tr[\sigma\widetilde\Lambda]
=\tr[\rho\Lambda]
\geq1-\eps.
\ee
Convexity of the squared Hilbert--Schmidt norm gives
\be
\tr[\widetilde\Lambda^2]
\leq\sum_z t_z\tr[(U_z^*\Lambda U_z)^2]
=\tr[\Lambda^2].
\ee
Thus the minimum quadratic cost for $\sigma$ cannot exceed
that for $\rho$. Taking ceilings and logarithms proves
\eqref{quadratic-schur}.

For the upper bound in \eqref{entropy-comparison},
every feasible projector in \eqref{smooth-max-spectral}
is a feasible effect with integer quadratic cost
$\tr[\Pi^2]=\tr[\Pi]$.
The ceiling of the minimum quadratic cost therefore
cannot exceed the minimum feasible projector rank.

Nonnegativity and the zero-error value were established
above. Increasing $\eps$ enlarges the feasible set, so
the quadratic entropy is nonincreasing.
For a pure state, its zero-error value is zero;
nonnegativity and monotonicity then imply that it
vanishes for every allowed error.
The corresponding properties of $H_{\max}^\eps$
follow directly from \eqref{smooth-max-spectral}.
\end{proof}

Like the trace-distance smoothed max-entropy, its
quadratically smoothed counterpart admits an explicit
spectral expression. The following lemma evaluates it
using the optimizer of the quadratic program
in~\cite[Sec.~III]{AR2020}.

\bmyl
\label{thm:quadratic-spectral}
Let $\rho\in\md(A)$ have positive eigenvalues
$p_1\geq\cdots\geq p_r>0$, and let $\eps\in(0,1)$.
For $\ell\in\{0,\ldots,r-1\}$, put
$S_\ell\eqdef\sum_{x=\ell+1}^r p_x^2$.
There is a unique $\ell\in\{0,\ldots,r-1\}$ satisfying
$\|\rho\|_{(\ell)}<1-\eps$ and
\be\label{quadratic-threshold}
p_\ell
\geq\frac{S_\ell}{1-\eps-\|\rho\|_{(\ell)}}
>p_{\ell+1},
\ee
where $p_0\eqdef\infty$ and $\|\rho\|_{(0)}\eqdef0$.
For this $\ell$,
\be\label{quadratic-spectral}
H_{\max}^{\eps,{\rm q}}(A)_\rho
=\log\left\lceil
\ell+\frac{\left(1-\eps-\|\rho\|_{(\ell)}\right)^2}{S_\ell}
\right\rceil.
\ee
\emyl

\begin{proof}
To evaluate the entropy, we first determine the minimum
quadratic cost. Pinching an effect in an eigenbasis of $\rho$
preserves its acceptance probability and cannot increase
its quadratic cost. We may therefore restrict to diagonal
effects whose weights outside $\supp(\rho)$ are zero.
The resulting spectral optimization has the known optimal
weights $\min\{1,cp_x\}$~\cite[Sec.~III]{AR2020}, where $c>0$
satisfies
\be\label{quadratic-normalization}
\sum_{x=1}^r p_x\min\{1,cp_x\}=1-\eps\,.
\ee
The left-hand side is continuous, equals zero at $c=0$,
and increases strictly to one as $c$ increases to $1/p_r$.
Thus $\eps\in(0,1)$ determines $c$ uniquely.

To compute the cost, we identify the weights equal to one.
Let $\ell$ be the number of eigenvalues satisfying
$cp_x\geq1$, including equality. Since $\eps>0$, at least
one weight is smaller than one, so
$\ell\in\{0,\ldots,r-1\}$. Splitting
\eqref{quadratic-normalization} at this index gives
\be
1-\eps=\|\rho\|_{(\ell)}+cS_\ell,
\qquad
\frac1c=\frac{S_\ell}{1-\eps-\|\rho\|_{(\ell)}}.
\ee
The denominator is positive, and the definition of $\ell$
gives
\be
p_\ell\geq\frac1c>p_{\ell+1}.
\ee
Thus $\ell$ satisfies \eqref{quadratic-threshold}, proving
existence.

For uniqueness, any index satisfying the stated conditions
defines a positive scalar
$c=(1-\eps-\|\rho\|_{(\ell)})/S_\ell$.
The threshold inequalities ensure that the weights
$\min\{1,cp_x\}$ equal one on the first $\ell$ eigenvalues
and equal $cp_x$ on the remainder. Their acceptance
probability is $1-\eps$, so this scalar satisfies
\eqref{quadratic-normalization} and must equal its unique
solution. The strict right inequality then makes $\ell$
exactly the number of eigenvalues at least $1/c$.
Hence $\ell$ is unique.

Finally, the optimal effect $\Lambda_*=\min\{I,c\rho\}$ (minimum understood spectrally)
has quadratic cost
\be
\tr[\Lambda_*^2]
=\ell+c^2S_\ell
=\ell+\frac{\big(1-\eps-\|\rho\|_{(\ell)}\big)^2}{S_\ell}.
\ee
Applying the ceiling and logarithm proves
\eqref{quadratic-spectral}.
\end{proof}

We have now compared the two smoothings and expressed both
in terms of the source spectrum. To connect them with
compression, we must show that each optimization describes
both an attainable protocol and a bound on every allowed
protocol. Theorems~\ref{thm:lo} and~\ref{thm:locc} establish
these identifications for LO and LOCC, respectively.
We begin with LO, where the projector formulation suggests
encoding the eigenspace carrying the largest source weight.

\section{Quantum Compression under LO}\label{sec:lo}

The projector characterization of $H_{\max}^\eps$ suggests
encoding the eigenspace carrying the largest source weight.
To prove that this prescription gives the compression cost,
we must also rule out an improvement from more general channels.
We therefore begin with arbitrary encoders and decoders. Let
$C_x:A\to A'$ and $D_y:A'\to A$ be their Kraus operators,
where $|A'|=m$ and the output is identified with $A$.
Equation~\eqref{ex14} then gives
\be\label{1p19}
F_\rho^2\left(\id_m\xrightarrow{\rm LO}\id_{A\to B}\right)
=\max_{\mE,\mD}\sum_{x,y}|\tr[D_yC_x\rho]|^2.
\ee
The next theorem shows that this unrestricted optimization
indeed reduces to the spectral subspace suggested by the entropy.

\bmyt\label{thm:lo}
For $\rho\in\md(A)$, $d\eqdef|A|$, and $m\in[d]$,
\be\label{ineqs}
F_\rho\left(\id_m\xrightarrow{\rm LO}\id_{A\to B}\right)
=\|\rho\|_{(m)}.
\ee
Consequently, for every $\eps\in[0,1)$,
\be\label{lbmax}
\comp_{\rm LO}^\eps(\rho)=H_{\max}^\eps(A)_\rho.
\ee
\emyt
\begin{proof}
Identify $B$ with $A$, and let $A'$ be an $m$-dimensional system.
We use~\eqref{1p19}
with $\{C_x:A\to A'\}$ and $\{D_y:A'\to A\}$ denoting Kraus
operators of the encoding and decoding channels, respectively.
For the lower bound, write
$\rho=\sum_{z\in[d]}p_z|z\lr z|$, with the eigenvalues in
decreasing order. Let $V:A'\to A$ be the isometry defined by
$V|z\ra_{A'}=|z\ra_A$ for $z\in[m]$.
Take $\mD$ to be the isometry channel induced by $V$, and take
$\mE$ to have Kraus operators $\{C_x\}_{x=0}^{d-m}$, where
$C_0\eqdef V^*$ and
$C_x\eqdef|1\ra_{A'}{}_A\la m+x|$ for $x\in[d-m]$.
When $m=d$, only $C_0$ is present. These operators satisfy
$\sum_{x=0}^{d-m}C_x^*C_x=I_A$. Moreover,
$\tr[VC_x\rho]=0$ for $x\in[d-m]$, whereas
\be
\tr[VC_0\rho]
=\sum_{z\in[m]}p_z
=\|\rho\|_{(m)}\;.
\ee
Hence~\eqref{1p19} gives
$F_\rho^2(\id_m\xrightarrow{\rm LO}\id_{A\to B})
\geq\|\rho\|_{(m)}^2$.

For the upper bound, fix $\mE$ and $\mD$. We show that, for every
$X\in\ml(A)$,
\be\label{eq:bottleneck-bound}
\Big(\sum_{x,y}|\tr[D_yC_xX]|^2\Big)^{1/2}
\leq\sum_{z\in[m]}s_z(X)\;,
\ee
where $s_z(X)$ are the singular values of $X$ in decreasing order.
Applying this to $X=\rho$ will give the desired upper bound
in~\eqref{1p19}.

By Euclidean norm duality,
\be
\Big(\sum_{x,y}|\tr[D_yC_xX]|^2\Big)^{1/2}
=\max_{\|\Lambda\|_2\leq1}
\Big|\sum_{x,y}\overline{\lambda_{xy}}\tr[D_yC_xX]\Big|\;,
\ee
where $\Lambda=(\lambda_{xy})$ and
$\|\Lambda\|_2^2=\sum_{x,y}|\lambda_{xy}|^2$.
Thus, setting
$K_\Lambda\eqdef\sum_{x,y}\overline{\lambda_{xy}}D_yC_x$,
we obtain
\ba\label{pre}
\Big(\sum_{x,y}|\tr[D_yC_xX]|^2\Big)^{1/2}
&=\max_{\|\Lambda\|_2\leq1}|\tr[K_\Lambda X]|\\
&\leq\max_{\|\Lambda\|_2\leq1}
\sum_{z\in[d]}s_z(K_\Lambda)s_z(X)\;,
\ea
where the inequality is von Neumann's trace inequality.

Write $s_x\eqdef s_x(X)$ and
$t_x\eqdef s_x(K_\Lambda)$, with singular values in
nonincreasing order. To bound the right-hand side of~\eqref{pre}
by $\sum_{x=1}^m s_x$, it suffices to show that for all $x\in[d]$
\be\label{eq:K-Lambda-norm-bounds}
0\leq t_x\leq1\qquad\text{and}\qquad
\sum_{x\in[d]} t_x\leq m\,.
\ee
Indeed, these constraints give
\ba
\sum_{x\in[d]} t_xs_x
&\leq\sum_{x\in[m]} t_xs_x
+s_m\sum_{x=m+1}^d t_x\\
&\leq\sum_{x\in[m]}
\big[t_xs_x+(1-t_x)s_m\big]\\
&\leq\sum_{x=1}^m s_x.
\ea
Here the first inequality uses $s_x\leq s_m$ for $x>m$,
the second uses the total weight bound in~\eqref{eq:K-Lambda-norm-bounds}, and the last
uses $s_m\leq s_x$ and $t_x\leq1$ for $x\leq m$.

We now establish \eqref{eq:K-Lambda-norm-bounds}.
For unit vectors $|\psi\ra,|\phi\ra\in A$,
Cauchy--Schwarz gives
\ba
|\la\phi|K_\Lambda|\psi\ra|^2
&\leq\sum_{x,y}|\la\phi|D_yC_x|\psi\ra|^2\\
&=\la\phi|(\mD\circ\mE)(|\psi\lr\psi|)|\phi\ra\\
&\leq1.
\ea
Maximizing over these vectors gives $t_1\leq1$,
and hence $0\leq t_x\leq1$ for every $x$.

For the total weight, choose a unitary $U\in\mfu(d)$
such that $\tr[UK_\Lambda]=\sum_x t_x$;
such a unitary exists by polar decomposition.
In an orthonormal basis $\{|z\ra\}_{z\in[m]}$ of $A'$,
cyclicity of the trace and Cauchy--Schwarz give
\ba\label{eq:trace-K-Lambda}
|\tr[UK_\Lambda]|
&\leq\sum_{z\in[m]}
\left|\sum_{x,y}\overline{\lambda_{xy}}
\la z|C_xUD_y|z\ra\right|\\
&\leq\sum_{z\in[m]}
\left(\sum_{x,y}|\la z|C_xUD_y|z\ra|^2\right)^{1/2}.
\ea
Each summand is at most one, since
\ba
\sum_{x,y}|\la z|C_xUD_y|z\ra|^2
&\leq\sum_{x,y}
\la z|D_y^*U^*C_x^*C_xUD_y|z\ra\\
&=1,
\ea
where we used the completeness relations for the encoder
and decoder and the unitarity of $U$.
Therefore $\sum_x t_x\leq m$.
This proves \eqref{eq:K-Lambda-norm-bounds} and hence
\eqref{eq:bottleneck-bound}.
Applying the latter to $X=\rho$
and using~\eqref{1p19} gives
$F_\rho^2(\id_m\xrightarrow{\rm LO}\id_{A\to B})
\leq\|\rho\|_{(m)}^2$.
Together with the lower bound, this proves the fidelity formula.
The cost identity follows from \eqref{defc} and
\eqref{smooth-max-spectral}.
\end{proof}

The converse is what makes the entropic characterization exact:
every pair of local channels is bounded by the same source weight
that spectral truncation attains. In particular, a fixed isometric
decoder suffices under LO. We next ask what changes when the
decoder may depend on a message from Alice.

\section{Quantum Compression under LOCC}\label{sec:locc}

We now allow Alice and Bob to use local operations and exchange
classical messages freely, without preshared entanglement.
A general LOCC superchannel converting a channel from $A'$
to $B'$ into one from $A$ to $B$ can be written as
\be\label{locc-general}
\Theta[\mN]
=\mD_{A''B'\to B}\circ
\mN_{A'\to B'}\circ
\mE_{A\to A''A'},
\ee
where $\mE$ is a local channel at Alice's laboratory,
$A''$ is a quantum memory retained by Alice, and $\mD$
is an LOCC channel acting on Alice's $A''$ and Bob's $B'$.
This form also accommodates classical exchanges before
transmission: their record can be stored in $A''$, and
any conditional ancillary state at Bob's laboratory can
be prepared during postprocessing.

For the noiseless resource $\id_m$, this description
simplifies considerably. Every channel attainable by such
a protocol can also be implemented with a classical memory
$Z$ in place of $A''$: Alice performs an instrument and
sends its quantum output through $\id_m$, together with
a free classical message specifying the outcome. Bob then
applies the corresponding isometry. Thus, classical
communication allows his decoder to depend on Alice's
encoding branch, while quantum memory at Alice and
messages from Bob to Alice do not enlarge the set of
attainable channels. The following lemma establishes this
reduction. It is a channel-resource formulation of the
characterization in~\cite{N2013}; we include a direct proof
to make its application to compression explicit.

\bmyl
\label{lem:locc-reduction}
Let $\mM\in\cptp(A\to B)$ and $m\leq|B|$.
The following statements are equivalent:
\begin{enumerate}
\item $\mM$ can be implemented using one transmission through
$\id_m$ and LOCC, without preshared entanglement.
Equivalently, $\mM=\Theta[\id_m]$ for an LOCC superchannel
$\Theta$ of the form~\eqref{locc-general}.
\item $\mM$ admits a Kraus representation $\{M_z\}_z$
with $\operatorname{rank}(M_z)\leq m$ for every $z$.
\item $\mM$ can be implemented by an instrument at Alice,
followed by transmission through $\id_m$ and a conditional
isometry at Bob. Explicitly,
\be\label{conditional-isometry-form}
\mM(\omega)
=\sum_z V_zC_z\omega C_z^*V_z^*,
\ee
where $C_z:A\to\mathbb C^m$,
$\sum_zC_z^*C_z=I_A$, and each
$V_z:\mathbb C^m\to B$ is an isometry.
\end{enumerate}
The same equivalence holds if $\id_m$ is replaced by any
unitary channel between $m$-dimensional systems.
\emyl

\begin{proof}
To reduce an LOCC implementation to the form in
\eqref{conditional-isometry-form}, we first show that its
Kraus operators have rank at most $m$. We then factor each
operator into an encoding branch and an isometric decoder.

Consider an implementation as in the first statement.
Alice's preprocessing has Kraus operators
$E_x:A\to A''A'$, where $|A'|=m$ and $A''$ is her
retained memory. Since Bob has no input or preshared
entanglement, his preparatory operations can be deferred
to the postprocessing, conditional on the classical
record of the preprocessing.

After the transmission, the LOCC postprocessing admits
a product Kraus representation. Since Alice has no final
output, refining the Kraus representation to include the
discarding of her remaining systems gives operators of
the form $\la a_y|_{A''}\otimes D_y$, where $D_y:B'\to B$.
Identifying $A'$ and $B'$ through $\id_m$, the resulting
channel therefore has Kraus operators
\be
M_{y,x}
=D_y\big(\la a_y|_{A''}\otimes I_{A'}\big)E_x.
\ee
Each operator factors through an $m$-dimensional system,
so its rank is at most $m$. This proves the second statement.

Conversely, suppose $\{M_z\}_z$ is such a Kraus
representation. To implement its $z$th branch, choose an isometry $V_z:\mathbb C^m\to B$ whose range
contains the range of $M_z$, and put $C_z\eqdef V_z^*M_z$.
Since $V_zV_z^*$ is the projector onto the range of $V_z$,
it fixes every vector in the range of $M_z$. Hence
\be
V_zC_z=V_zV_z^*M_z=M_z.
\ee
Thus, $\{C_z\}_z$ defines an instrument since
\ba
V_zC_z&=M_z,\\
\sum_zC_z^*C_z&=\sum_zM_z^*M_z=I_A\,.
\ea
Alice sends its
quantum output through $\id_m$ and communicates $z$ to
Bob, who applies $V_z$. This implements $\mM$ and proves
the third statement. The first statement follows immediately.

Finally, let the resource be the unitary channel
$\mU(\omega)=U\omega U^*$. Replacing each encoding
operator $C_z$ by $U^*C_z$ gives the same implementation.
Conversely, $\mU$ can be implemented using $\id_m$ followed
by the local unitary $U$.
\end{proof}

The lemma turns the optimization over LOCC protocols into
an optimization over Kraus representations: the communication
dimension restricts the rank of each Kraus operator.
Combining this characterization with the source-fidelity
formula in~\eqref{ex14}, we obtain
\be\label{mzlo}
F_\rho^2\left(
\id_m\xrightarrow{\rm LOCC}\id_{A\to B}
\right)
=
\max_{\substack{\sum_zM_z^*M_z=I_A\\
\operatorname{rank}(M_z)\leq m}}
\sum_z\big|\tr[M_z\rho]\big|^2,
\ee
where $B$ is identified with $A$. Every feasible Kraus
representation is attainable by the protocol in the lemma,
so this expression accounts for both achievability and
the converse. The remaining optimization depends only on the source spectrum.
The next theorem evaluates it and identifies the quadratic
entropy as its compression cost.

The quadratic optimization underlying
$H_{\max}^{\eps,{\rm q}}$ was previously obtained for
partially smoothed conditional min-entropy~\cite{AR2020}.
Its connection to compression already gives a converse:
teleporting the quantum message turns a classically assisted
compression protocol into a state-merging protocol.
The state-merging converse~\cite[Theorem~6]{ABJT2020},
combined with the quadratic reduction in~\cite{AR2020}
and the corresponding error conversion, therefore yields
\be
\comp_{\rm LOCC}^\eps(\rho)
\geq H_{\max}^{\eps,{\rm q}}(A)_\rho\,.
\ee

The following theorem shows that this bound is attained
exactly and identifies the optimal fidelity for every
communication dimension. The main step is an explicit
compression protocol attaining the quadratic optimum.
We also give a direct converse proof, so that the operational
characterization is established within the present framework.

\bmyt\label{thm:locc}
Let $\rho\in\md(A)$, $d\eqdef|A|$, and $m\in[d]$. Then
\be\label{locc-effect-form}
F_\rho\left(\id_m\xrightarrow{\rm LOCC}\id_{A\to B}\right)
=\max_{\substack{0\leq\Lambda\leq I\\
\tr[\Lambda^2]\leq m}}\tr[\rho\Lambda].
\ee
Consequently, for every $\eps\in[0,1)$,
\be\label{comp-quadratic}
\comp_{\rm LOCC}^\eps(\rho)
=H_{\max}^{\eps,{\rm q}}(A)_\rho\,.
\ee
\emyt

\noindent\textbf{Remarks.}
\ben
\item Lemma~\ref{thm:quadratic-spectral} gives, for $\eps\in(0,1)$,
\be\label{locc-cost-spectral}
\comp_{\rm LOCC}^\eps(\rho)
=\log\left\lceil
\ell+\frac{\left(1-\eps-\|\rho\|_{(\ell)}\right)^2}{S_\ell}
\right\rceil,
\ee
where $\ell$ and $S_\ell$ are defined in that lemma.
At zero error, the cost is $\log|\supp(\rho)|$.

\item The effect optimization in \eqref{locc-effect-form}
is the $m$-distillation norm evaluated on the eigenvalue
vector of $\rho$. Its known spectral evaluation
\cite[Theorem~2]{RFWA2018} gives
\be\label{loccexact}
F_\rho\left(\id_m\xrightarrow{\rm LOCC}\id_{A\to B}\right)
=\|\rho\|_{(k)}+(m-k)b_k
\ee
when $m<r\eqdef\rank(\rho)$, and equals one
otherwise. Here $b_j\eqdef\sqrt{S_j/(m-j)}$ for
$j\in\{0,\ldots,m-1\}$, $S_j$ is defined in
Lemma~\ref{thm:quadratic-spectral}, and $k$ is the unique index satisfying
\be\label{loccthreshold}
p_k\geq b_k>p_{k+1}\,,
\ee
with the conventions
$p_0\eqdef\infty$, and $\|\rho\|_{(0)}\eqdef0$. Thus $k$ is determined by the communication dimension $m$,
whereas $\ell$ in Lemma~\ref{thm:quadratic-spectral}
is determined by the error $\eps$.
\een

\begin{proof}
Let $s$ denote the right-hand side of
\eqref{locc-effect-form}. We first evaluate $s$, then
give a direct proof that it bounds the LOCC source fidelity.
By \eqref{mzlo}, this requires showing that every admissible
Kraus representation satisfies
\be
\sum_z|\tr[M_z\rho]|^2\leq s^2.
\ee
We then construct a Kraus representation attaining equality,
which Lemma~\ref{lem:locc-reduction} converts into an
LOCC compression protocol.

We may assume $m<\rank(\rho)$, since otherwise
exact transmission and the support projector attain one
on the two sides of~\eqref{locc-effect-form}, respectively.
Restricting to $\supp(\rho)$, we may further assume
$\rho>0$, so that $m<d$. Indeed, projecting the output
onto this support and replacing its orthogonal complement
by a state in the support fixes the target and cannot
decrease fidelity. Conversely, a protocol on the support
extends to the full input space. Restricting effects to
the support likewise preserves their objective and cannot
increase their quadratic cost.

To compute $s$, pinch the effect in an eigenbasis of $\rho$.
This preserves its objective and cannot increase its quadratic
cost, reducing the optimization to maximizing
$\sum_xp_x\lambda_x$ subject to $0\leq\lambda_x\leq1$
and $\sum_x\lambda_x^2\leq m$, where
$p_1\geq\cdots\geq p_d$ are the eigenvalues of $\rho$
in nonincreasing order.
Without the upper bounds $\lambda_x\leq1$, Cauchy--Schwarz
makes weights proportional to $p_x$ optimal. This suggests
the clipped candidate $\mu_x\eqdef\min\{1,p_x/b\}$.
We choose its scale $b$ to exhaust the quadratic budget:
\be\label{locc:cutoff}
\sum_{x\in[d]}\min\{1,p_x^2/b^2\}=m\,.
\ee
Viewed as a function of $b$, the sum on the left-hand side is continuous and
equals $d$ for $0<b\leq p_d$. As $b$ increases beyond
$p_d$, it decreases strictly from $d$ towards zero.
Since $0<m<d$, equation~\eqref{locc:cutoff} therefore
has a unique solution.

Let $k$ be the number of saturated weights, corresponding
to the eigenvalues $p_x\geq b$. Thus
$p_k\geq b>p_{k+1}$, as in \eqref{loccthreshold}. Equation~\eqref{locc:cutoff} implies $k<m$
and gives
\be\label{mk}
\sum_{x=k+1}^d p_x^2=(m-k)b^2.
\ee
Thus $b=b_k$ and $p_k\geq b_k>p_{k+1}$.
Conversely, any index satisfying \eqref{loccthreshold}
gives a solution $b_k$ of \eqref{locc:cutoff}.
Uniqueness of that solution forces $b_k=b$, so the
threshold condition determines $k$ uniquely as the
number of eigenvalues at least $b$.

To verify optimality of the candidate, Cauchy--Schwarz gives,
for every feasible $\{\lambda_x\}_x$,
\be
\sum_x\mu_x\lambda_x
\leq\sqrt{\sum_x\mu_x^2}\sqrt{\sum_x\lambda_x^2}
\leq m\,.
\ee
Since $\mu_x=1$ for $x\leq k$ and $b\mu_x=p_x$
for $x>k$, every feasible $\{\lambda_x\}_x$ satisfies
\ba
\sum_xp_x\lambda_x
&=b\sum_x\mu_x\lambda_x
  +\sum_{x\in[k]}(p_x-b)\lambda_x\\
&\leq bm+\sum_{x\in[k]}(p_x-b)\\
&=\|\rho\|_{(k)}+(m-k)b.
\ea
The inequality also uses $\lambda_x\leq1$ and
$p_x\geq b$ for $x\leq k$.
Equality holds at $\lambda_x=\mu_x$, proving that
$s=\|\rho\|_{(k)}+(m-k)b$.

We next show that $s$ also bounds the LOCC source fidelity.
For this, it suffices to construct $P>0$ with
$\tr[P]=s^2$ such that
\be\label{loccwitness}
|\tr[M\rho]|^2\leq\tr[PM^*M]
\ee
for every operator $M$ of rank at most $m$.
Indeed, summing over an admissible Kraus representation gives
\be\label{s2}
\sum_z|\tr[M_z\rho]|^2
\leq\tr\left[P\sum_zM_z^*M_z\right]
=s^2\,.
\ee

The key to proving~\eqref{loccwitness} is to write
$M\rho=XY$, where $X\eqdef MP^{1/2}$ and
$Y\eqdef P^{-1/2}\rho$. Since $\operatorname{rank}(X)\leq m$,
the singular-value trace inequality followed by
Cauchy--Schwarz gives
\ba\label{xy}
|\tr[XY]|
&\leq\sum_{x=1}^m s_x(X)s_x(Y)\\
&\leq\|X\|_2
\Big(\sum_{x\in[m]} s_x(Y)^2\Big)^{1/2}.
\ea
Thus, to obtain \eqref{loccwitness}, we seek $P$ with
$\tr[P]=s^2$ for which $\sum_{x\in[m]} s_x(Y)^2=1$.
In an eigenbasis of $\rho$, choose
\be\label{locc:witness-operator}
P\eqdef
s\sum_{x=1}^k p_x|x\lr x|
+\frac{s}{b}\sum_{x=k+1}^d p_x^2|x\lr x|.
\ee
Since $\sum_{x=k+1}^d p_x^2=(m-k)b^2$, we have
\be
\tr[P]
=s\left(\|\rho\|_{(k)}+(m-k)b\right)=s^2.
\ee
Moreover,
\be
\rho P^{-1}\rho
=\frac1s\sum_{x=1}^k p_x|x\lr x|
+\frac{b}{s}\sum_{x=k+1}^d|x\lr x|.
\ee
By \eqref{loccthreshold}, its first $k$ eigenvalues are
at least $b/s$, and the remaining $d-k$ eigenvalues
equal $b/s$. There are therefore at least $m-k$ tail
eigenvalues, giving
\ba\label{locc:singular-sum}
\sum_{x=1}^m s_x(Y)^2
&=\|\rho P^{-1}\rho\|_{(m)}\\
&=\frac{\|\rho\|_{(k)}+(m-k)b}{s}\\
&=1.
\ea
Substituting into~\eqref{xy} and squaring yields
\ba
|\tr[M\rho]|^2
&\leq\|MP^{1/2}\|_2^2\\
&=\tr[P^{1/2}M^*MP^{1/2}]\\
&=\tr[PM^*M].
\ea
This proves~\eqref{loccwitness}, and hence the converse
bound~\eqref{s2}.

To attain the bound, let $J\eqdef\{k+1,\ldots,d\}$,
$t_x\eqdef p_x^2/b^2$ for $x\in J$, and let
$\bt\in\mathbb{R}^J$ be the vector with components $t_x$.
The threshold condition $p_k\geq b>p_{k+1}$ implies
$0<t_x<1$, while \eqref{mk} gives
$\sum_{x\in J}t_x=n$, where $n\eqdef m-k$.
Thus $\bt$ belongs to the polytope
\be
\mP\eqdef\left\{
\br\in\mathbb{R}^J\;:\;
0\leq r_x\leq1\,,\quad \sum_{x\in J}r_x=n
\right\}\,.
\ee
We will express $\bt$ as a convex combination of the
vertices of $\mP$.
These vertices are precisely the vectors with $n$ coordinates
equal to one and all others equal to zero. \footnote{Indeed, any
nonintegral point has at least two fractional coordinates,
since its coordinate sum is an integer. Perturbing these
coordinates in opposite directions shows that such a point
is not a vertex. Conversely, every zero-one vector in
$\mP$ is a vertex.}

To index these vertices, let $\mi_n$ denote the collection
of all $n$-element subsets of $J$; it contains
$\binom{d-k}{n}$ subsets. Each $T\in\mi_n$ corresponds
to the incidence vector $\mathbf{1}_T\in\mathbb{R}^J$,
defined by
\be
(\mathbf{1}_T)_x\eqdef
\begin{cases}
1 &\text{if }x\in T\\
0 &\text{if }x\notin T
\end{cases}
\qquad x\in J.
\ee
Since $\bt\in\mP$, there is therefore a probability
distribution $\{w_T\}_{T\in\mi_n}$ such that
\be
\bt=\sum_{T\in\mi_n}w_T\mathbf{1}_T\,.
\ee
For each $x\in J$, let
$\mi_{n,x}\eqdef\{T\in\mi_n:\,x\in T\}$
denote the subsets containing $x$.
Taking the $x$th coordinate of this convex combination gives
\be\label{tt}
\sum_{T\in\mi_{n,x}}w_T=t_x\,.
\ee

We now construct Kraus operators attaining the bound.
For each $T\in\mi_n$, define
\be
M_T\eqdef\sqrt{w_T}\left(
\sum_{x\in[k]}|x\ra\la x|
+\sum_{x\in T}\frac{b}{p_x}|x\ra\la x|
\right).
\ee
Each operator has rank at most $k+n=m$.
We now verify that $\sum_{T\in\mi_n}M_T^*M_T=I$
and that these operators attain the bound.
Exchanging the order of summation gives
\ba
\sum_{T\in\mi_n}M_T^*M_T
&=\sum_{x\in[k]}|x\ra\la x|
+\sum_{x\in J}\frac1{t_x}
 \left(\sum_{T\in\mi_{n,x}}w_T\right)|x\ra\la x|\\
&=\sum_{x\in[k]}|x\ra\la x|
+\sum_{x\in J}|x\ra\la x|\\
&=I,
\ea
where the second equality follows from~\eqref{tt}.
Moreover,
\be
\tr[M_T\rho]
=\sqrt{w_T}\left(\|\rho\|_{(k)}+nb\right)
=\sqrt{w_T}\,s,
\ee
and hence
\be
\sum_{T\in\mi_n}|\tr[M_T\rho]|^2
=s^2\sum_{T\in\mi_n}w_T=s^2.
\ee
Thus the Kraus optimization attains $s^2$, proving
\eqref{locc-effect-form} and~\eqref{loccexact}.

Finally, a dimension $m$ meets the required fidelity if and only
if some effect satisfies $\tr[\rho\Lambda]\geq1-\eps$ and
$\tr[\Lambda^2]\leq m$. The smallest such integer dimension
is the ceiling of the minimum quadratic cost, proving
\eqref{comp-quadratic}. The spectral cost follows from
Lemma~\ref{thm:quadratic-spectral}; at zero error, the
minimum dimension is $|\supp(\rho)|$.
\end{proof}

The spectral expression in \eqref{loccexact} equals the
$m$-distillation norm of the eigenvalue vector $(p_x)_{x=1}^r$.
This norm and its spectral evaluation were introduced in
one-shot coherence distillation~\cite[Theorem~2]{RFWA2018}
and subsequently used in pure-state entanglement
distillation~\cite[Sec.~3.2]{RFWG2019}.
In those settings, it acts on basis amplitudes or Schmidt
amplitudes; here it acts on the source eigenvalues themselves.
Theorem~\ref{thm:locc} establishes its operational meaning
for compression, with an attaining protocol and a matching
converse that preserve the inaccessible reference.

The attaining protocol also requires only a finite classical
message. Suppose $m<r\eqdef\operatorname{rank}\rho$, and
let $k$ be the threshold index in \eqref{loccthreshold}.
Working on $\supp(\rho)$, the vector $\bt$ has $r-k$
coordinates whose sum is fixed at $n=m-k$.
Its polytope therefore has affine dimension at most
$r-k-1$. By Carath\'eodory's theorem, $\bt$ can be expressed
as a convex combination of at most $r-k$ incidence vectors.
Thus the weights $w_T$ can be chosen with at most $r-k$
nonzero entries, giving an attaining instrument with at
most $r-k$ outcomes of nonzero probability on $\rho$.
Alice can communicate its outcome using at most
$\lceil\log_2(r-k)\rceil$ bits.
This bounds the classical communication required by the
constructed protocol; its minimum possible cost is not
determined here.

\subsection{The advantage of classical communication}

We can now compare the operational models directly. The LO
formula retains only the first $m$ eigenvalues, whereas the
LOCC formula incorporates the entire positive tail through
its quadratic contribution. This contribution is strictly
larger whenever nonzero source weight lies beyond the first
$m$ eigenvalues.

\bmyc\label{cor:advantage}
Let $\rho\in\md(A)$ and $m<\operatorname{rank}\rho$.
Then
\be\label{loccadvantage}
F_\rho\left(\id_m\xrightarrow{\rm LOCC}\id_{A\to B}\right)
>
F_\rho\left(\id_m\xrightarrow{\rm LO}\id_{A\to B}\right).
\ee
Moreover, for every $\eps$ satisfying
\be\label{advantage-errors}
1-F_\rho\left(\id_m\xrightarrow{\rm LOCC}\id_{A\to B}\right)
\leq\eps<1-\|\rho\|_{(m)},
\ee
the compression costs obey
\be
\comp_{\rm LOCC}^\eps(\rho)
\leq\log(m)
<
\comp_{\rm LO}^\eps(\rho).
\ee
The interval in \eqref{advantage-errors} is nonempty.
\emyc

\begin{proof}
With $k$ as in \eqref{loccthreshold}, Cauchy--Schwarz gives
\be
\sum_{x=k+1}^m p_x
\leq\sqrt{(m-k)\sum_{x=k+1}^m p_x^2}
<\sqrt{(m-k)\sum_{x=k+1}^d p_x^2}.
\ee
The last inequality is strict because
$m<\operatorname{rank}\rho$. Adding $\|\rho\|_{(k)}$
and using the exact fidelity formulas proves
\eqref{loccadvantage}, so the stated interval is nonempty.
For every error in this interval, LOCC attains the
required fidelity using dimension $m$, whereas LO
requires a larger dimension.
\end{proof}

The distinction is already present at $m=1$, when no quantum
system is transmitted: the squared fidelities under LO and
LOCC are $p_1^2$ and $\tr[\rho^2]$, respectively. Consequently,
\be
\comp_{\rm LOCC}^\eps(\rho)=0<
\comp_{\rm LO}^\eps(\rho)
\qquad\text{whenever}\qquad
1-\|\rho\|_2\leq\eps<1-\|\rho\|_\infty\,.
\ee
This interval is nonempty for every mixed source. At these
thresholds, classical communication alone suffices to meet
the fidelity requirement, whereas LO requires quantum
transmission.

For a flat spectrum on a support of size $r$, the squared
fidelities are $(m/r)^2$ under LO and $m/r$ under LOCC,
for $m\leq r$. The latter agrees with the Schmidt-number
benchmark for process fidelity~\cite{NT2012}, since the
source purification is maximally entangled on that support.
For example, a maximally mixed qubit source at
$\eps=1-1/\sqrt2$ requires dimension two under LO but
only dimension one under LOCC.

Both models require dimension $\rank(\rho)$
at zero error, and their costs can also coincide at positive
error because dimensions are integers. Nevertheless, the
examples above exhibit a strict saving. We next show that
this saving has no uniform bound over sources, even at a
fixed error threshold.

\bmyc\label{cor:unbounded-advantage}
Fix $\eps\in(0,1)$. For an integer $n\geq(1+\eps)/(1-\eps)$, let $\rho_n$ have rank $n^2+1$
and eigenvalues
\be\label{unbounded-spectrum}
p_1\eqdef1-\frac{n\eps}{n-1},
\qquad
p_x\eqdef\frac{\eps}{n(n-1)},
\quad x=2,\ldots,n^2+1.
\ee
Then
\be\label{unbounded-costs}
\comp_{\rm LO}^\eps(\rho_n)=\log_2(n+1),
\qquad
\comp_{\rm LOCC}^\eps(\rho_n)=1.
\ee
Consequently, at every fixed $\eps\in(0,1)$, the single-shot
compression advantage is unbounded as the source rank increases.
\emyc

\begin{proof}
We show that fidelity $1-\eps$ requires dimension $n+1$
under LO, whereas dimension two is necessary and sufficient
under LOCC. The spectrum is chosen so that adding $n$ tail
eigenvalues to $p_1$ gives exactly $1-\eps$.
The eigenvalues are normalized since $p_1+n^2p_2=1$.
The condition on $n$ ensures $p_1\geq np_2$, and
$p_2>0$, so they are positive and ordered.
By \eqref{ineqs}, the optimal LO fidelity is the sum
of the largest eigenvalues. Since
$p_1+np_2=1-\eps$, exactly $n$ tail eigenvalues must
accompany $p_1$ to reach the target.
Thus the minimum LO dimension is $n+1$.

For LOCC, dimension two suffices. Indeed, at $m=2$,
the choice $k=1$ gives
\be
b_1=\sqrt{\sum_{x=2}^{n^2+1}p_x^2}=np_2.
\ee
This threshold satisfies \eqref{loccthreshold} because
$p_1\geq np_2$ and $np_2>p_2$. Hence
\be
F_{\rho_n}\left(\id_2\xrightarrow{\rm LOCC}\id_{A\to B}\right)
=p_1+np_2=1-\eps.
\ee
Dimension one is insufficient, since
\ba
F_{\rho_n}\left(\id_1\xrightarrow{\rm LOCC}\id_{A\to B}\right)
&=\sqrt{p_1^2+n^2p_2^2}\\
&<p_1+np_2\\
&=1-\eps,
\ea
where strictness follows from positivity.
Thus the minimum LOCC dimension is two,
proving \eqref{unbounded-costs}.
\end{proof}

For example, at $\eps=0.01$, taking $n=101$ gives a source
of rank $10202$ whose minimum compression dimensions are
$102$ under LO and two under LOCC. The unbounded saving
above is obtained by varying the source. We next consider independent and identically distributed
(i.i.d.) sources, asking how the saving behaves when
compressing increasingly many copies of one fixed source.

\section{Asymptotic Compression of i.i.d. Sources}
\label{sec:asymptotics}

Throughout this section, logarithms are base two and
$\eps\in(0,1)$ is fixed. Let $p_1,\ldots,p_r$ be the positive
eigenvalues of $\rho$, counted with multiplicity, and let $X$
take the value $-\log(p_x)$ with probability $p_x$.
We call the log-spectrum nonlattice if there are no
$a\in\mathbb R$ and $h>0$ such that all these values
belong to $a+h\mathbb Z$. Write $H\eqdef\mathbb E[X]$,
$V\eqdef\mathbb E[(X-H)^2]$, and
$\kappa_3\eqdef\mathbb E[(X-H)^3]$. Let $\Phi$ 
denote the standard normal distribution function.
The exact single-shot formulas express the two compression
costs through spectral optimizations studied in classical
source coding and quantum information theory. We first
state the established asymptotic results in our conventions,
then determine the constant-order difference between LO
and LOCC when the log-spectrum is nonlattice.

\subsection{Established rates and converse results}

The quantum source-coding theorem identifies $H$ as the
asymptotic compression rate~\cite{S1995}. The strong converse
for general encoders and decoders~\cite[Chap.~I]{W1999}
also applies under the entanglement-fidelity criterion.
With free classical communication, the strong converse
follows from~\cite[Theorem~8]{Sh2014}: replacing the quantum
transmission by teleportation converts an LOCC compression
protocol into a state-merging protocol with trivial initial
quantum side information at Bob and the same fidelity.

More precisely, consider any sequence of protocols for
$\rho^{\otimes n}$ using the noiseless channel $\id_{m_n}$,
where $m_n$ is a positive integer. No fidelity requirement
is imposed on this sequence. For either model, the optimal
source fidelity tends to zero whenever
\be
\limsup_{n\to\infty}\frac{\log(m_n)}{n}<H.
\ee
Thus no such sequence can attain source fidelity at least
$1-\eps$ for any fixed $\eps\in(0,1)$. Together with the source-coding theorem, this gives, for every
fixed $\eps\in(0,1)$,
\be\label{loccasymptotic}
\lim_{n\to\infty}\frac1n
\comp_{\rm LOCC}^\eps(\rho^{\otimes n})
=
\lim_{n\to\infty}\frac1n
\comp_{\rm LO}^\eps(\rho^{\otimes n})
=H.
\ee

At second order, upper and lower bounds for blind compression
were obtained in~\cite[Theorem~5.8(ii)]{DL2015}.
The gap was closed in~\cite{AR2020} using the
partial-smoothing state-merging converse
of~\cite[Theorem~6]{ABJT2020}.
In our source-fidelity convention, the established LO
expansion is
\be\label{asym:second}
\comp_{\mf}^\eps(\rho^{\otimes n})
=nH+\sqrt{nV}\,\Phi^{-1}(1-\eps)
+O\big(\log(n)\big),
\qquad
\mf\in\{{\rm LO},{\rm LOCC}\}.
\ee
The same expansion holds for LOCC because
\eqref{comp-quadratic} identifies its cost with the rounded
quadratic program whose second-order asymptotics were
also determined in~\cite{AR2020}.
The purified-distance error in that analysis is
$\sqrt{1-(1-\eps)^2}$, giving the quantile
$\Phi^{-1}(1-\eps)$ above. For a mixed source, rounding
changes the logarithmic cost by $o(1)$; for a pure source,
both costs vanish.
Thus classical communication leaves the established
second-order term unchanged. Before examining the finer
terms that reveal its asymptotic advantage, we recall
the known fidelity-decay exponents below the entropy rate.

The strong converse describes what happens at rates below
$H$: the optimal fidelity tends to zero. Exponent formulas
quantify how rapidly this occurs. We recall the known LO exponent~\cite[Theorem~15]{BY2025}
and obtain the LOCC exponent by combining the pure-state
partial-smoothing exponent~\cite[Theorem~5]{BY2025}
with our exact characterization \eqref{locc-effect-form}.

Fix a communication rate $r\geq0$ and put
$m_n\eqdef\lfloor2^{nr}\rfloor$. Consistently with our
fidelity convention, define
\be\label{asym:exponent-definition}
E_{\mf}(r)
\eqdef\lim_{n\to\infty}-\frac1n\log\,
F_{\rho^{\otimes n}}
\left(\id_{m_n}\xrightarrow{\mf}\id_{A^n\to B^n}\right),
\ee
where $A^n$ and $B^n$ denote the $n$-copy input and output
systems. The results below establish that these limits
exist. Dimensions exceeding the input dimension permit
exact transmission. 

The LO exponent was determined
in~\cite[Theorem~15]{BY2025} for squared fidelity.
Its value in our convention is therefore divided by two.
For LOCC, let $\lambda_x^{(n)}$ denote the positive
eigenvalues of $\rho^{\otimes n}$. Pinching the effect
optimization in \eqref{locc-effect-form} gives
\ba
f_n
&\eqdef
F_{\rho^{\otimes n}}\!
\left(\id_{m_n}\xrightarrow{\rm LOCC}\id_{A^n\to B^n}\right)\\
&=
\max_{\substack{0\leq\mu_x\leq1\\
\sum_x\mu_x^2\leq m_n}}
\sum_x\lambda_x^{(n)}\mu_x.
\ea
This is the inverse tradeoff of the pure-state quadratic
program in~\cite[Sec.~III]{AR2020}: instead of minimizing
the quadratic cost at a prescribed fidelity, we maximize
the fidelity at a prescribed cost.

To connect this optimization with the known exponent,
let $\psi_{RA}$ be a purification of $\rho$, and put
$\psi_{R^nA^n}\eqdef\psi_{RA}^{\otimes n}$. Define
\be\label{asym:partial-smoothing-error}
\delta_n\eqdef
\min_{\tau_{R^nA^n}\geq0}
\sqrt{1-
\la\psi_{R^nA^n}|\tau_{R^nA^n}|\psi_{R^nA^n}\ra},
\ee
where the minimum is subject to
\be
\tau_{R^n}\leq\psi_R^{\otimes n},
\qquad
\tau_{R^nA^n}
\leq m_n\,\psi_R^{\otimes n}\otimes I_{A^n}.
\ee
Here $\tau_{R^n}\eqdef\tr_{A^n}[\tau_{R^nA^n}]$.
The marginal constraint ensures
$\tr[\tau_{R^nA^n}]\leq1$. Thus the objective is the
purified distance from the pure target to the
subnormalized operator $\tau_{R^nA^n}$.
These constraints define the partial-smoothing problem
with operator budget $m_n$.
The pure-state reduction in~\cite[Sec.~III]{AR2020}
identifies its maximum overlap with $f_n^2$. Hence
\be
f_n=\sqrt{1-\delta_n^2},
\qquad
\sqrt{1-\delta_n}
\leq f_n\leq\sqrt{2(1-\delta_n)}.
\ee
Consequently, the fidelity exponent is one-half of the
decay exponent of $1-\delta_n$ used in~\cite{BY2025}.

The integer rounding of the budget does not affect this
exponent. Indeed, let $\tilde f_n$ denote the same
quadratic optimum with budget $2^{nr}$ in place of $m_n$.
Scaling its feasible weights by $\sqrt{m_n/2^{nr}}$
makes them feasible for budget $m_n$. Since
$m_n\geq2^{nr}/2$, it follows that
\be
\frac{\tilde f_n}{\sqrt2}
\leq f_n\leq\tilde f_n.
\ee
The normalized negative logarithms therefore differ
by at most $1/(2n)$.

In~\cite[Theorem~5]{BY2025}, a partial-smoothing rate $s$
corresponds to operator budget $2^{-ns}$.
Our communication rate $r$ therefore corresponds to
$s=-r$. Applying that theorem and converting to our
fidelity convention gives
\be
E_{\rm LOCC}(r)
=\min_{\mathbf q}
\left\{D(\q\|\p)+\frac12[H(\q)-r]_+\right\}\,.
\ee
Here $\p$ is the positive eigenvalue vector of $\rho$,
and the minimum is over probability vectors $\q$ on
the same alphabet. We use the Shannon entropy
$H(\q)\eqdef-\sum_xq_x\log(q_x)$, the classical
relative entropy
$D(\q\|\p)\eqdef\sum_xq_x\log(q_x/p_x)$,
and $[a]_+\eqdef\max\{a,0\}$.

To express this formula in terms of R\'enyi entropies 
$H_\alpha(\rho)\eqdef(1-\alpha)^{-1}\log\tr[\rho^\alpha]$ (with $\alpha\ge 1$),
write
\be
\frac12[H(\q)-r]_+
=\max_{0\leq t\leq1/2}t\big(H(\q)-r\big).
\ee
The objective is convex in $\q$ and affine in $t$,
so minimax permits interchanging the minimum and maximum.
For fixed $t$, put $Z_t\eqdef\sum_y p_y^{1/(1-t)}$
and $q_x^\star\eqdef p_x^{1/(1-t)}/Z_t$. Then
\be
D(\mathbf q\|\p)+tH(\mathbf q)
=(1-t)D(\mathbf q\|\mathbf q^\star)
-(1-t)\log (Z_t)\,.
\ee
Nonnegativity of relative entropy shows that the unique
minimizer is $\mathbf q^\star$, with minimum value
$tH_{1/(1-t)}(\rho)$.
Substitution gives
$t\big(H_{1/(1-t)}(\rho)-r\big)$.
Setting $\alpha=1/(1-t)$ yields the LOCC expression
below. Together with the known LO exponent, we obtain
\ba\label{asym:known-exponents}
E_{\rm LO}(r)
&=\sup_{\alpha\geq1}\frac{\alpha-1}{\alpha}
\big(H_\alpha(\rho)-r\big),\\
E_{\rm LOCC}(r)
&=\max_{1\leq\alpha\leq2}\frac{\alpha-1}{\alpha}
\big(H_\alpha(\rho)-r\big),
\ea
where the objective at $\alpha=1$ is zero by continuity.
Both exponents are positive for $r<H$ and vanish for
$r\geq H$.

The LOCC fidelity exponent follows from our exact
single-shot characterization \eqref{locc-effect-form}
and the known pure-state partial-smoothing exponent
in~\cite[Theorem~5]{BY2025}, after translating its error
convention to ours.
For trivial initial quantum side information at Bob, the
one-way state-merging exponent
in~\cite[Theorem~3.1]{BCRT2026} also reduces to the same
R\'enyi expression after converting its squared-fidelity
convention to ours. That theorem concerns net entanglement
cost, whereas our derivation identifies this expression
with the fidelity exponent of compression without
preshared entanglement.
Related fixed-length pure-state coding exponents,
including the exact visible strong-converse exponent,
were established earlier under the ensemble-average
fidelity criterion~\cite{H2002}, which differs from
the preservation of a source purification required here.

The exponents above describe fidelity decay below the entropy
rate. In the complementary low-error moderate-deviation regime,
the purified-distance error tends to zero while the compression
rate approaches $H$. The corresponding expansion for quantum
source coding was established
in~\cite[Theorem~IV.3]{RTB2023}.
We now keep $\eps\in(0,1)$ fixed and determine the
constant-order compression saving provided by classical
communication beyond the common leading terms.

\subsection{The classical spectral benchmark}

By \eqref{lbmax}, the minimum LO compression dimension is
the smallest number of strings carrying probability at least
$1-\eps$ under the product distribution $\p^{\otimes n}$.
Its asymptotics therefore follow from classical fixed-length
source coding~\cite{S1964,H2008,KV2014}.
For LOCC, the clipped spectral optimizer of the quadratic
program in~\cite[Sec.~III]{AR2020}, together with standard
concentration estimates~\cite{KV2014,P1975}, yields the
same logarithmic term. Consequently, for $V>0$,
\be\label{asym:third}
\comp_{\mf}^\eps(\rho^{\otimes n})
=nH+\sqrt{nV}\Phi^{-1}(1-\eps)-\frac12\log(n)+O(1),
\qquad
\mf\in\{{\rm LO},{\rm LOCC}\}.
\ee
A third-order refinement for quantum compression was also
announced in~\cite{RSlides}. We record the common logarithmic
term as background; our comparison below concerns the
constant hidden in the $O(1)$ remainder.

For a nonlattice spectrum, classical source-coding asymptotics
also determine the LO constant; see, in particular,
Ref.~\cite[Sec.~V]{H2018}. In our base-two convention, put
\be\label{asym:constant}
C(\rho,\eps)
\eqdef\frac{\kappa_3}{6V}\left(\left(\Phi^{-1}(1-\eps)\right)^2-1\right)
+\log\left(\frac{\phi(z)}{\sqrt V\,\ln(2)}\right).
\ee
The known expansion then reads
\be\label{asym:LO-fourth}
\comp_{\rm LO}^\eps(\rho^{\otimes n})
=nH+\sqrt{nV}\Phi^{-1}(1-\eps)-\frac12\log(n)+C(\rho,\eps)+o(1).
\ee
We have stated this result for the logarithm of an arbitrary
integer compression dimension, as defined in \eqref{defc};
rounding the cost itself to an integer number of qubits would
introduce an additional bounded rounding term. We do not
reprove \eqref{asym:LO-fourth}.

\subsection{The constant-order advantage of classical communication}

The preceding results fix the unassisted benchmark and the
first three orders of the assisted cost. The exact LOCC
formula allows us to evaluate the remaining constant and
compare the required dimensions directly.

\bmyt[Constant-order comparison]\label{thm:higher-order}
Suppose $V>0$ and the spectrum of $\rho$ is nonlattice. Then
\be\label{asym:fourth}
\comp_{\rm LOCC}^\eps(\rho^{\otimes n})
=nH+\sqrt{nV}\Phi^{-1}(1-\eps)-\frac12\log(n)
+C(\rho,\eps)+1-\log_2(e)+o(1).
\ee
Consequently,
\be\label{asym:advantage}
\lim_{n\to\infty}
\left(
\comp_{\rm LO}^\eps(\rho^{\otimes n})
-\comp_{\rm LOCC}^\eps(\rho^{\otimes n})
\right)
=\log_2(e)-1.
\ee
\emyt
\begin{proof}
Let $M_{n,{\rm LO}}$ and $M_{n,{\rm LOCC}}$ denote the minimum
compression dimensions at fidelity at least $1-\eps$.
Since the LO expansion \eqref{asym:LO-fourth} is known,
it suffices to prove
\be
\frac{M_{n,{\rm LOCC}}}{M_{n,{\rm LO}}}
=\frac2e+o(1).
\ee
To compute the assisted dimension, we use its quadratic
characterization in \eqref{comp-quadratic}. The optimizer
is determined by a single spectral cutoff. We first find
the cutoff that meets the fidelity constraint and then
evaluate the quadratic cost at that cutoff.

We begin with the known LO benchmark. Put
\be
a_n\eqdef nH+\sqrt{nV}\,\Phi^{-1}(1-\eps)
+\frac{\kappa_3}{6V}
\left([\Phi^{-1}(1-\eps)]^2-1\right).
\ee
Exponentiating \eqref{asym:LO-fourth} gives
\be\label{asym:LO-size}
M_{n,{\rm LO}}
=\frac{2^{a_n}e^{-\frac12[\Phi^{-1}(1-\eps)]^2}}
{\ln(2)\,\sqrt{2\pi nV}}
\big(1+o(1)\big).
\ee
We will compare the assisted dimension with this expression.

By \eqref{comp-quadratic}, the assisted dimension is the
minimum quadratic cost $\tr[\Lambda^2]$, rounded upward,
over effects satisfying
$\tr[\rho^{\otimes n}\Lambda]\geq1-\eps$.
The optimizer has the known clipped form
$\Lambda(a)\eqdef\min\{I,2^a\rho^{\otimes n}\}$,
where the minimum is defined spectrally
\cite[Sec.~III]{AR2020}.
Thus an eigenvalue $\lambda_x$ of $\rho^{\otimes n}$
receives weight one when $\lambda_x>2^{-a}$ and weight
$2^a\lambda_x$ otherwise. The parameter $a$ is chosen
so that the fidelity constraint holds with equality.

To determine $a$ and the resulting cost, separate the
eigenvalues above the cutoff from those below it.
Let $S_n$ be a sum of $n$ independent copies of $X$.
Equivalently, $S_n$ takes the value $-\log(\lambda_x)$
with probability $\lambda_x$, with the positive eigenvalues
$\lambda_x$ counted with multiplicity. Define
\ba\label{asym:cutoff-functions}
A_n(a)
&\eqdef\sum_{\lambda_x>2^{-a}}\lambda_x
=\Pr\{S_n<a\},\\
N_n(a)
&\eqdef\big|\{x:\lambda_x>2^{-a}\}\big|
=\mathbb E\!\left[2^{S_n}\mathbf1_{\{S_n<a\}}\right],\\
B_n(a)
&\eqdef2^a\sum_{\lambda_x\leq2^{-a}}\lambda_x^2
=\mathbb E\!\left[2^{a-S_n}\mathbf1_{\{S_n\geq a\}}\right].
\ea
Here $A_n(a)$ is the total probability of the eigenvalues
retained fully, and $N_n(a)$ counts them.
The remaining eigenvalues contribute $B_n(a)$ to the
fidelity constraint and $2^aB_n(a)$ to the quadratic cost.
Consequently, the optimal cutoff $b_n$ satisfies
\ba\label{asym:LOCC-dimension}
A_n(b_n)+B_n(b_n)&=1-\eps,\\
M_{n,{\rm LOCC}}
&=\left\lceil N_n(b_n)+2^{b_n}B_n(b_n)\right\rceil.
\ea
The first equation has a unique solution because its
left-hand side is continuous and strictly increasing
until it reaches one.

We now estimate these quantities near $a_n$.
The nonlattice one-term Edgeworth expansion
\cite{P1975,H1992} and the nonlattice local limit theorem
\cite{S1965} give, uniformly for $t$ in any fixed bounded
interval,
\ba\label{asym:local-estimates}
A_n(a_n+t)
&=1-\eps
+\frac{e^{-\frac12[\Phi^{-1}(1-\eps)]^2}}
{\sqrt{2\pi nV}}\,t
+o(n^{-1/2}),\\
2^{-(a_n+t)}N_n(a_n+t)
&=\frac{e^{-\frac12[\Phi^{-1}(1-\eps)]^2}}
{\ln2\,\sqrt{2\pi nV}}
+o(n^{-1/2}),\\
B_n(a_n+t)
&=\frac{e^{-\frac12[\Phi^{-1}(1-\eps)]^2}}
{\ln2\,\sqrt{2\pi nV}}
+o(n^{-1/2}).
\ea
The correction in $a_n$ cancels the first Edgeworth
correction, giving the first line.
For the last two lines, the local limit theorem estimates
interval probabilities on either side of the cutoff.
Weighting these probabilities by $2^{-y}$ gives the
coefficient $\int_0^\infty2^{-y}\,dy=1/\ln2$.
A uniform $O(n^{-1/2})$ bound for probabilities of
fixed-length intervals controls the remaining weighted
tails. This argument applies to discrete nonlattice
distributions and does not assume that $S_n$ has a density.
Finite support supplies the required moments; no additional
Cram\'er condition is needed.

We first use these estimates to locate $b_n$.
Adding the first and third lines gives
\be
A_n(a_n+t)+B_n(a_n+t)
=1-\eps
+\frac{e^{-\frac12[\Phi^{-1}(1-\eps)]^2}}
{\sqrt{2\pi nV}}
\left(t+\frac1{\ln2}\right)
+o(n^{-1/2}).
\ee
The fidelity constraint therefore suggests
$t=-1/\ln2$.
To justify this, fix $\delta>0$ and evaluate the preceding
expression at $t=-1/\ln2-\delta$ and
$t=-1/\ln2+\delta$.
For all sufficiently large $n$, the two values lie below
and above $1-\eps$, respectively. Monotonicity gives
\be
a_n-\frac1{\ln2}-\delta
<b_n<
a_n-\frac1{\ln2}+\delta.
\ee
Since $\delta$ is arbitrary,
$b_n=a_n-1/\ln2+o(1)$.

We can now evaluate the quadratic cost at $b_n$.
Since $b_n-a_n$ remains bounded, the uniform estimates
in \eqref{asym:local-estimates} apply at this cutoff.
Their last two lines give
\ba\label{asym:LOCC-size}
N_n(b_n)+2^{b_n}B_n(b_n)
&=\frac{2^{b_n+1}e^{-\frac12[\Phi^{-1}(1-\eps)]^2}}
{\ln2\,\sqrt{2\pi nV}}
\big(1+o(1)\big)\\
&=\frac2e\,
\frac{2^{a_n}e^{-\frac12[\Phi^{-1}(1-\eps)]^2}}
{\ln2\,\sqrt{2\pi nV}}
\big(1+o(1)\big).
\ea
In the second line we used
$2^{b_n-a_n}=e^{-1}(1+o(1))$.
Since $V>0$ implies $H>0$, this cost grows exponentially.
Rounding it upward therefore changes it by a relative
$o(1)$.
Comparing \eqref{asym:LOCC-size} with
\eqref{asym:LO-size} proves the desired dimension ratio.
Taking logarithms gives \eqref{asym:advantage}, and combining
this comparison with \eqref{asym:LO-fourth} proves
\eqref{asym:fourth}.
\end{proof}

The comparison also gives
$M_{n,{\rm LOCC}}/M_{n,{\rm LO}}\to2/e$.
Thus classical communication saves asymptotically
$\log_2(e)-1\approx0.442695$ bits per block, while the
saving per source copy vanishes. In natural logarithmic
units, the same saving is $1-\ln2$ nats per block.

The nonlattice assumption enters at constant order.
For lattice spectra with $V>0$, the common expansion
\eqref{asym:third} remains valid, but its bounded correction
can depend on the position of the cutoff within the
lattice~\cite{KV2014,H2018}. The constant-order limit
\eqref{asym:advantage} is therefore asserted only for
nonlattice spectra. In particular, every nonuniform
full-rank qubit spectrum belongs to the lattice case.

When $V=0$, the source is uniform on its support, and the
exact single-shot formulas give a different constant-order
comparison. Writing $r\eqdef\operatorname{rank}\rho$, for
$r\geq2$ we obtain
\ba
\comp_{\rm LO}^\eps(\rho^{\otimes n})
&=\log\left\lceil r^n(1-\eps)\right\rceil,\\
\comp_{\rm LOCC}^\eps(\rho^{\otimes n})
&=\log\left\lceil r^n(1-\eps)^2\right\rceil.
\ea
Since $1-\eps>0$ is fixed, integer rounding contributes
$o(1)$ to these logarithms, and the saving converges to
$-\log(1-\eps)$ bits per block.
For $r=1$, the source is pure and both costs are zero.

\section{Discussion}\label{sec:discussion}

Taken together, Theorems~\ref{thm:lo} and~\ref{thm:locc}
identify the operational role of classical communication
in approximate quantum compression. An optimal LO protocol
uses one fixed decoding subspace, and its fidelity is the
source weight contained in that subspace. Classical
communication allows the decoding subspace to depend on
the encoder's outcome. The attaining construction shows
how these branches use spectral weight beyond any one
$m$-dimensional subspace, while the converse proves that
no LOCC protocol can exceed their performance.
Corollary~\ref{cor:advantage} makes the distinction precise:
classical communication strictly improves fidelity whenever
the transmitted dimension is smaller than the source rank.
A reduction in compression cost additionally depends on
the error threshold, because the available dimensions are
integers.

The entropic identities \eqref{lbmax} and
\eqref{comp-quadratic} express this operational distinction
through two smoothings of the same zero-error quantity.
Both models require the log-rank at zero error; assistance
changes how approximation can reduce that cost.
The projector optimization becomes a quadratic effect
optimization, and Theorem~\ref{thm:locc} establishes that
this relaxation describes an attainable communication
protocol. This exact identification is useful beyond the
comparison itself: together with the spectral formulas,
it gives a benchmark for evaluating finite-block protocols
and determining precisely when a classical message saves
quantum communication.

The size of this saving depends on how the source is varied.
Corollary~\ref{cor:unbounded-advantage} shows that, at every
fixed positive error, the saving has no uniform bound over
source states. Theorem~\ref{thm:higher-order} addresses a
different question: repeated copies of one fixed source.
For positive entropy variance and a nonlattice spectrum,
classical communication leaves the established first three
orders unchanged but reduces the constant term by
$\log_2(e)-1$. Although the saving per copy vanishes,
the minimum communication dimensions have a nontrivial
limiting ratio of $2/e$. Thus the leading entropy rate
does not capture the full resource comparison, and the
exact single-shot formulas reveal a distinction that
persists at constant order.

A central direction is to determine whether the exact
characterizations obtained here extend to quantum state
merging. In that task, Alice transfers $A$ to Bob, who
already holds $B$, while preserving the joint purification
$\psi_{RAB}$. The resources include classical communication
and shared entanglement. Our LOCC compression problem
corresponds to the special case of trivial initial side
information at Bob, initial maximally entangled rank $m$,
and no required final entanglement: the two tasks interconvert
by teleportation and by transmitting half of a maximally
entangled state. Theorem~\ref{thm:locc} therefore provides
an exactly solved case from which to investigate the
general problem.

The concrete question is whether one can determine the
optimal merging fidelity at prescribed initial and final
entanglement ranks, with a matching converse and an attaining
protocol. Partially smoothed conditional min-entropy already
gives one-shot bounds~\cite[Theorem~6]{ABJT2020}, and the
asymptotic strong-converse exponent is known~\cite{BCRT2026}.
An exact finite-resource characterization would go further:
it would determine how the correlations available to Bob
affect the full fidelity--resource tradeoff. The quadratic
optimization in this paper suggests a starting point, but
quantum side information introduces operator structure that
cannot generally be reduced to the eigenvalues of $\rho_A$.
A key question is whether the partial-smoothing converse
can be made exactly attainable, or whether additional
constraints are needed to describe physical merging
protocols.

This leads to a broader research program: identifying
conditional entropic optimizations that characterize quantum
information transfer exactly at finite resources. State
merging is a natural first test, with state splitting and
state redistribution providing further settings in which
quantum communication, classical communication, and
entanglement can be varied separately. The present results
demonstrate the usefulness of such exact identifications:
they connect an entropic optimization to explicit protocols
and reveal resource advantages hidden by the leading
asymptotic rate. Establishing analogous connections for
correlated sources would substantially extend this picture.

\section*{Acknowledgments}
This research was supported by the Israel Science Foundation
under Grant No.~1192/24.\\

\noindent\textbf{Use of AI.} The main results, including Theorems 5 and 7, were established more than a year ago. At that stage, the authors used AI only for editorial assistance and discussion. More recently, the authors used GPT-6 Astra via Codex for substantial assistance in exploring, developing, and checking the i.i.d. asymptotic analysis leading to Theorem~\ref{thm:higher-order}, as well as in preparing the manuscript and organizing references. The authors independently verified the mathematical arguments and references and take responsibility for all claims and the final manuscript.

\appendix

\bibliography{references}

\section{Extensions of Operator Smoothing}\label{app:smoothing}

The compression entropies share an acceptance constraint but
assign different costs to the accepted effect. We first extend
this comparison by varying the effect cost, then describe how
admissible effects can select states on which a general
divergence is smoothed.

\subsection{Varying the effect cost}

For $\rho\in\md(A)$ and $\eps\in[0,1)$, define
\be\label{effect-smoothing-set}
\mT_\eps(\rho)
\eqdef\{0\leq\Lambda\leq I:\ \tr[\rho\Lambda]\geq1-\eps\}.
\ee
The projector characterization \eqref{smooth-max-spectral}
equivalently gives
\be
H_{\max}^\eps(A)_\rho
=\log\min_{\Lambda\in\mT_\eps(\rho)}
\operatorname{rank}\Lambda,
\ee
because replacing an effect by its support projector preserves
its rank and increases its acceptance probability.

More genegrally, let $g:[0,1]\to[0,1]$ be nondecreasing
and lower semicontinuous, with $g(0)=0$, $g(1)=1$,
and $g(t)>0$ for $t>0$. Define
\be\label{general-effect-entropy}
H_g^\eps(A)_\rho
\eqdef\log\min_{\Lambda\in\mT_\eps(\rho)}
\tr[g(\Lambda)].
\ee
The indicator cost $g(t)=\mathbf1_{\{t>0\}}$ recovers
$H_{\max}^\eps$. For $g(t)=t^2$, this is the unrounded
quadratic quantity: the compression entropy
$H_{\max}^{\eps,{\rm q}}(A)_\rho$ equals
$\log\lceil2^{H_g^\eps(A)_\rho}\rceil$.
Every such $H_g^\eps$ reduces to $\log|\supp(\rho)|$
at zero error, since a feasible effect then acts as the
identity on $\supp(\rho)$, and the support projector
attains the minimum.

For $g(t)=t$, \eqref{general-effect-entropy} equals
$-D_H^\eps(\rho\|I)$, where we use the hypothesis-testing
convention~\cite{WR2012}
\be
D_H^\eps(\rho\|\sigma)
\eqdef-\log\min_{\Lambda\in\mT_\eps(\rho)}
\tr[\sigma\Lambda].
\ee
In particular, the minimum is not divided by $1-\eps$.
For general $g$ and $\sigma\geq0$, define
\be\label{relative-effect-cost}
D_g^\eps(\rho\|\sigma)
\eqdef-\log\min_{\Lambda\in\mT_\eps(\rho)}
\tr[\sigma g(\Lambda)].
\ee
At zero error, this equals
$D_{\min}(\rho\|\sigma)\eqdef-\log\tr[\Pi_\rho\sigma]$,
where $\Pi_\rho$ is the support projector~\cite{D2009}.
Indeed, feasibility forces $g(\Lambda)\geq\Pi_\rho$,
with equality attained by $\Lambda=\Pi_\rho$.

If $g$ is operator convex, $D_g^\eps$ satisfies data
processing. An effect $\Lambda$ admissible for $\mN(\rho)$
pulls back to $\mN^*(\Lambda)\in\mT_\eps(\rho)$, and operator
Jensen's inequality~\cite{HP2003} gives
\ba
\tr[\sigma g(\mN^*(\Lambda))]
&\leq\tr[\sigma\mN^*(g(\Lambda))]\\
&=\tr[\mN(\sigma)g(\Lambda)].
\ea
Minimizing and applying $-\log$ yields
\be
D_g^\eps(\rho\|\sigma)
\geq D_g^\eps\big(\mN(\rho)\big\|\mN(\sigma)\big).
\ee
This includes $g(t)=t^\alpha$ for $1\leq\alpha\leq2$.
Thus the linear and quadratic costs belong to a family
of smoothings of $D_{\min}$.

Optimization over high-acceptance effects appears in
hypothesis testing~\cite{WR2012} and operator-smoothed
quasi-entropies~\cite[Sec.~VI.B.2]{BD2010}.
The latter construction inserts an effect into a
quasi-entropy; it does not establish the quadratic
compression identity \eqref{comp-quadratic}.
Here the effect formulation connects the two exact
compression characterizations.

\subsection{Smoothing a general divergence by domination}

Varying $g$ above changes the smoothing of $D_{\min}$.
To smooth a general divergence, we instead use an admissible
effect to select its first state argument. For
$\Lambda\in\mT_\eps(\rho)$, define
\be\label{effect-selected-state}
\rho_\Lambda
\eqdef
\frac{\sqrt{\rho}\,\Lambda\sqrt{\rho}}
{\tr[\rho\Lambda]}.
\ee
The numerator is bounded above by $\rho$ and has trace
at least $1-\eps$.

\bmyd
Let $D_f$ be a quantum $f$-divergence satisfying data
processing. For $\rho,\sigma\in\md(A)$ and $\eps\in[0,1)$,
define
\be\label{dominated-divergence-definition}
D_f^{\eps,{\rm dom}}(\rho\|\sigma)
\eqdef
\inf_{\Lambda\in\mT_\eps(\rho)}
D_f(\rho_\Lambda\|\sigma).
\ee
\emyd

The construction applies equally to any divergence satisfying
data processing, including quantum $f$-divergences defined
through integral and hockey-stick
representations~\cite{HT2024,HMT2026}.
Its channel compatibility follows from an equivalent
description of the admissible states.

\bmyt\label{dominated-smoothing-theorem}
For $\rho,\sigma\in\md(A)$ and $\eps\in[0,1)$,
\be\label{dominated-divergence-state-form}
D_f^{\eps,{\rm dom}}(\rho\|\sigma)
=
\inf_{\substack{\trho\in\md(A)\\
(1-\eps)\trho\leq\rho}}
D_f(\trho\|\sigma).
\ee
This quantity is nonincreasing in $\eps$, equals $D_f$
at zero error, and satisfies
\be\label{dominated-divergence-dpi}
D_f^{\eps,{\rm dom}}
\big(\mN(\rho)\big\|\mN(\sigma)\big)
\leq D_f^{\eps,{\rm dom}}(\rho\|\sigma)
\ee
for every $\mN\in\cptp(A\to B)$.
\emyt

\begin{proof}
Put $c\eqdef1-\eps$. If $\Lambda$ is admissible and
$a\eqdef\tr[\rho\Lambda]$, then
\be
c\rho_\Lambda
\leq a\rho_\Lambda
=\sqrt{\rho}\,\Lambda\sqrt{\rho}
\leq\rho.
\ee
Conversely, if $c\trho\leq\rho$, the operator
$\Lambda\eqdef c\rho^{-1/2}\trho\,\rho^{-1/2}$
is an effect satisfying $\tr[\rho\Lambda]=c$ and
$\rho_\Lambda=\trho$. Here the inverse is taken on
$\supp(\rho)$, and $\Lambda$ is zero on its orthogonal
complement. This proves
\eqref{dominated-divergence-state-form}.

The feasible sets increase with $\eps$, and at zero error
their only state is $\rho$, proving the first two properties.
For data processing, positivity of $\mN$ preserves
$c\trho\leq\rho$, so every feasible $\trho$ satisfies
\ba
D_f^{\eps,{\rm dom}}
\big(\mN(\rho)\big\|\mN(\sigma)\big)
&\leq D_f\big(\mN(\trho)\big\|\mN(\sigma)\big)\\
&\leq D_f(\trho\|\sigma).
\ea
Taking the infimum proves \eqref{dominated-divergence-dpi}.
\end{proof}

The domination constraint means that $\rho$ contains the
selected state with weight at least $1-\eps$.
For $\eps>0$, it is equivalent to
$\rho=(1-\eps)\trho+\eps\omega$ for some $\omega\in\md(A)$.
Hence
\be\label{dominated-smoothing-trace-distance}
\frac12\|\rho-\trho\|_1
=\frac{\eps}{2}\|\omega-\trho\|_1
\leq\eps.
\ee
Thus every admissible state belongs to the trace-distance
ball $\mb^\eps(\rho)$.
Equivalently, admissibility is the one-sided constraint
$D_{\max}(\trho\|\rho)\leq-\log(1-\eps)$, where~\cite{D2009}
\be
D_{\max}(\omega\|\rho)
\eqdef\log\inf\{\lambda>0:\ \omega\leq\lambda\rho\}.
\ee

Although domination is stronger than trace-distance
closeness, it gives the same smoothing of log-rank.

\bmyt\label{dominated-log-rank-theorem}
For $\rho\in\md(A)$ and $\eps\in[0,1)$,
\be\label{dominated-log-rank}
H_{\max}^\eps(A)_\rho
=
\min_{\substack{\trho\in\md(A)\\
(1-\eps)\trho\leq\rho}}
\log|\supp(\trho)|.
\ee
\emyt

\begin{proof}
The support projector $\Pi$ of every feasible $\trho$
satisfies $\tr[\rho\Pi]\geq1-\eps$.
The projector characterization \eqref{smooth-max-spectral}
therefore gives the lower bound.

Conversely, choose a minimum-rank spectral projector $\Pi$
satisfying $\tr[\rho\Pi]\geq1-\eps$, and put
$a\eqdef\tr[\rho\Pi]$. The normalized truncation
$\trho\eqdef\Pi\rho\Pi/a$ has log-rank
$H_{\max}^\eps(A)_\rho$ and satisfies
\be
(1-\eps)\trho
=\frac{1-\eps}{a}\Pi\rho\Pi
\leq\rho.
\ee
This proves the reverse bound.
\end{proof}

Domination smoothing thus recovers the LO entropy by selecting
a normalized state of minimum rank. The LOCC optimization
instead retains the quadratic cost of the selecting effect.
For a pure source, domination leaves the selected state
unchanged, whereas the minimum quadratic effect cost is
$(1-\eps)^2$. Its unrounded logarithm is $2\log(1-\eps)$;
the rounded entropy $H_{\max}^{\eps,{\rm q}}$ remains zero.
The state construction therefore does not replace quadratic
effect smoothing. Their relative versions also have opposite
error dependence: domination smoothing is nonincreasing in
$\eps$, while \eqref{relative-effect-cost} is nondecreasing.

\end{document}